\documentclass[12pt]{article}

\usepackage[english]{babel}
\usepackage[latin1]{inputenc}
\usepackage{graphicx}       
\usepackage{listings}
\usepackage{a4}
\usepackage{tikz,pgfplots}
\usepackage{inputenc}
\usepackage{subcaption}
\usepackage{listings}
\usepackage{epsfig}
\usepackage{amsmath,amssymb,amsfonts} 
\usepackage{relsize}
\usepackage{mathrsfs}
\usepackage{pstricks}
\usepackage{multirow}
\usepackage[all]{xypic}
\usepackage{colortbl}
\usepackage{rotating}
\usepackage{float}
\usepackage{fancyhdr}
\usepackage{ae}
\usepackage{natbib}
\usepackage{multicol}
\usepackage{enumerate}
\usepackage{wrapfig}
\usepackage[hang,flushmargin]{footmisc}
\usepackage{bold-extra}

\begin{document}

\begin{center}
{\noindent{\LARGE \textbf{A Bayesian model for lithology/fluid class prediction using a Markov mesh prior
fitted from a training image\vspace{1cm} \\} 
{\large {\large \textsc{H\aa kon Tjelmeland}}, \textsc{Xin Luo}} and 
{\large \textsc{Torstein Fjeldstad}}\\[0.2cm]{\large \it Department of
    Mathematical Sciences, Norwegian University of Science and
    Technology}}}
\vspace{0.4cm}
\end{center}

\begin{abstract}
We consider a Bayesian model for inversion of observed  amplitude variation with offset (AVO) data into lithology/fluid classes, and 
study in particular how the choice of prior distribution for the lithology/fluid classes influences
the inversion results. Two distinct prior distributions are considered, a simple manually specified 
Markov random field prior with a first order neighborhood and a Markov mesh model 
with a much larger neighborhood estimated from a training image. They are chosen to model both horisontal connectivity and 
vertical thickness distribution of the lithology/fluid classes,
and are compared on an offshore clastic oil reservoir in the North Sea. We combine
both priors with the same linearised Gaussian likelihood function based on a convolved linearised Zoeppritz relation and estimate properties
of the resulting two posterior distributions by simulating from these distributions
with the Metropolis--Hastings algorithm.

The influence of the prior on the marginal posterior probabilities for the lithology/fluid classes
is clearly observable, but modest. The importance of the prior on the connectivity properties
in the posterior realisations, however, is much stronger. The larger neighborhood of the 
Markov mesh prior enables it to identify and model connectivity and curvature much better than what can be 
done by the first order neighborhood Markov random field prior. As a result, we conclude that the posterior realisations based on the Markov mesh prior appear with much higher lateral connectivity, which is geologically plausible.

\end{abstract}

\vspace{0.5cm}
\noindent {\it Key words: Bayesian lithology/fluid class prediction, Markov chain Monte Carlo, 
  Markov mesh model, Metropolis--Hastings algorithm, profile Markov random field, 
  training image.} 
\vspace{-0.1cm}

\renewcommand{\baselinestretch}{1.00}   

\newcommand*{\plotnode}[2]{%
    \newcommand*{#1}{}
    \pgfmathsetmacro{#1}{#2}%
}%

\section{Introduction}

From seismic data one can predict elastic properties and lithology/fluid
classes (LFC) in a reservoir. This is an inverse problem  and for a given
seismic data set many solutions exist. Different methods have been used
for inverting seismic data to elastic properties and lithology/fluid 
classes, both deterministic approaches such as 
optimisation-based methods \citep{book41,book42} 
and probabilistic approaches such as Bayesian inversion \citep{book43}.
Using the Bayesian framework, a linearised relation between the data and 
the elastic properties is commonly used and a Gaussian likelihood function 
is adopted, see for example \citet{art162}, \citet{art163} and the discussion 
in \citet{art164}. When inverting to elastic properties the prior is also often 
assumed to be Gaussian, in which case the posterior becomes Gaussian with 
analytically available mean and covariance, see again \citet{art162}. 
When inverting to lithology/fluid classes, other priors have to be used. 
In particular \citet{art171} define a hierarchical prior, where a
Markov random field \citep{book28,col6}
is used to model the lithology/fluid classes and conditional on these 
the elastic properties are assumed to be Gaussian with mean and covariance functions
depending on the lithology/fluid classes. \citet{art174} consider a Gaussian mixture prior for the elastic attributes to include multimodality and skewness in the prior model where the effect of the lithology/fluid classes is summed out.
 With a non-Gaussian prior
the posterior is no longer analytically available and
Markov chain Monte Carlo \citep{book17,book24,book30} must typically 
be used to estimate
properties of the resulting posterior distribution. It is also challenging to specify lateral connectivity and spatial dependency laterally for non-Gaussian priors, and often the inverse problem is solved trace-by-trace before a smoother is applied afterwards \citep{art176}.

To specify a prior that reflects available prior information in a spatial problem 
like inversion of seismic data can be difficult. The properties of a Gaussian field is
analytically well understood, so by adopting a Gaussian prior as in \citet{art162}
the prior specification process is simplified. \citet{art172} consider a one
dimensional problem and assume a Markov chain prior to predict geological 
attributes from well log data, and \citet{tech34} use a similar model
to predict lithology/fluid classes and elastic attributes from seismic data. 
The properties of Markov chains are also analytically available, which 
again simplifies the specification of a reasonable prior. For most 
non-Gaussian spatial prior models the situation is less favorable. In 
 \citet{art146} a discrete Markov random field prior is used for the lithology/fluid
classes. The properties of discrete Markov random fields are analytically not 
available, which makes it difficult to verify the properties of the chosen prior.
To cope with the problem of specifying non-Gaussian spatial prior models it has
in geostatistics become common practice to estimate the prior model from a 
so-called training image. A training image is from an ou{tcrop or a constructed scene
assumed to have the same spatial structure as the phenomenon under study. The idea
is to estimate a prior model from one or more training images, 
see the discussion in \citet{book44}. Various multiple-point statistics models
\citep{pro23,art121,art125,art167} have been defined to implement this idea.
These models are algorithmically defined. The nodes in a lattice are visited
in a random order and when a node is visited, the value in that node is simulated 
conditional on values in previously visited nodes, where the conditional 
distribution used is estimated from the training image. There are two 
serious complications associated with the use of multiple-point statistics
models. First, the number of conditional distributions that has to be estimated from 
the training image is enormous, and the information content in a typical training 
image is not sufficient to estimate this number of parameters. \citet{art165} are
discussing this issue mathematically. Second, the models are only algorithmically 
defined and no simple to evaluate expressions are available for the estimated model. 
The implication of this is that if we want to use the estimated model as a prior
and generate realisations conditional on some observed data, it is in 
general not clear how to do this. Since we have no analytical formula for the 
prior, we neither have an expression for the posterior. This issue is also 
discussed in \citet{art168}. As alternatives to the multiple-point statistics models 
\citet{art157} and \citet{tech35} introduce procedures for fitting Markov random 
fields and Markov mesh models, respectively, to a given training image. For 
these model classes explicit expressions for the distributions are available, 
so to simulate from a corresponding conditional distribution Markov chain 
Monte Carlo procedures can for example be employed.

The purpose of this article is to demonstrate how inversion of seismic data 
into lithology/fluid classes can be accomplished in a Bayesian framework by estimating 
a prior model for the lithology/fluid classes from a given 
training image, and combine this with a linearised and Gaussian 
likelihood function. We fit a Markov mesh prior model to a training image
as discussed in \citet{tech35} and use Markov chain Monte Carlo to 
simulate from the resulting posterior distribution as discussed in 
\citet{art146}. A Markov mesh prior is used for lithology/fluid prediction
also in \citet{art175}, but they specify manually the neighborhood and interaction 
structures and fit only the parameter values to the training image. In our
fitting procedure we fit both the neighborhood and interaction structures and 
the parameter values to the given training image.
To focus on the methodological aspects we consider a
situation with only two lithology/fluid classes, oil sand and shale. 
In particular we compare the results from our procedure with what 
we get by instead using a simpler manually chosen Markov random 
field prior.  

The article has the following layout: First we present the data set and the associated training image, and we analyse and introduce 
our Bayesian model formulation. A Gaussian likelihood function is defined, and its properties are discussed. We introduce the Markov mesh and Markov random field priors, and in specific focus on how we fit the 
Markov mesh prior to the given training image. Next, a sampling algorithm for the posterior distribution is discussed. The two priors are tested on a real 2D section case study in the North Sea. Finally, we discuss the posterior properties of the two priors and provide some closing remarks.
 \section{Methodology}
The objective is to demonstrate and compare two different prior models in a Bayesian framework 
to predict lithology/fluid classes in the subsurface. In this section, we introduce the data 
set and formulate the inverse problem in a Bayesian setting, define a likelihood function and the two priors,
and discuss posterior simulation. 

\subsection{Data set and Bayesian model formulation}
\label{sec:data}
In this article we consider a seismic section from the Alvheim field in the North Sea, which 
is a clastic oil reservoir. The Alvheim field is characterised by a complex sand lobe geometry and is buried approximately $2$ km below the sea floor. In the analysis we
use one near and one far offset seismic data represented in a $105\times 51$ lattice 
$G=\{(i,j)|i=1,\ldots,105;j=1,\ldots,51\}$.  The stacked sections were generated from pre-stack time migrated (PSTM) common depth gathers (CDP), see \citet{art146} for futher processing details. The seismic data are shown in Figure \ref{fig:data}. 
The horizontal and vertical resolutions are about $100$ m and $4$ ms, respectively.
We let $d_{ij},(i,j)\in G$ denote a vector of size two containing the observed near and far offset seismic data
in node $(i,j)\in G$, and let $d$ be a vector where all $d_{ij},(i,j)\in G$ are stacked on top of each other.
\begin{figure}
\begin{center}
\begin{tabular}{@{}c@{}c@{}c@{}c@{}}
~~~& \makebox[6.5cm]{\includegraphics[width=6cm,angle=-90]{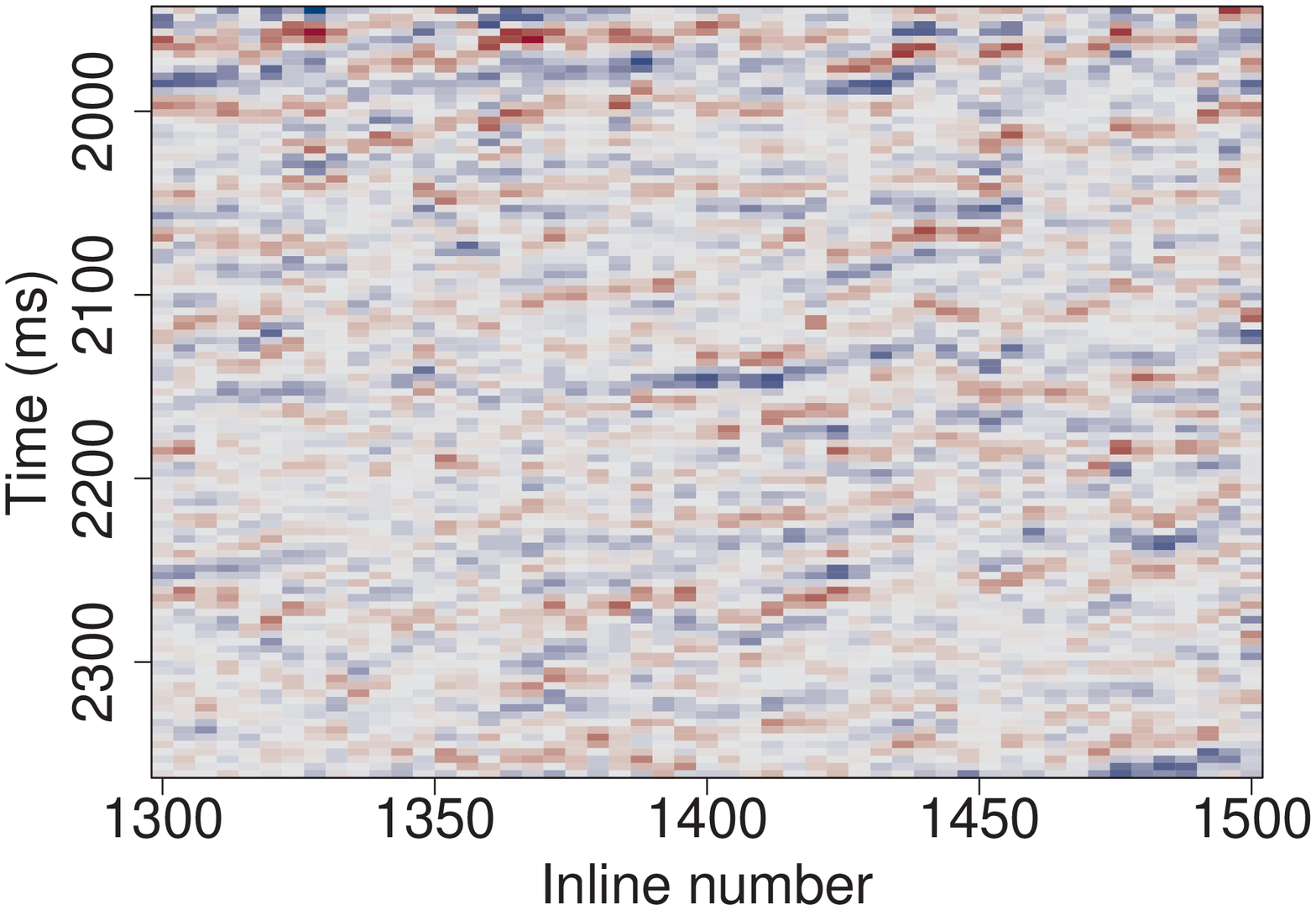}} &
\makebox[6.5cm]{\includegraphics[width=6cm,angle=-90]{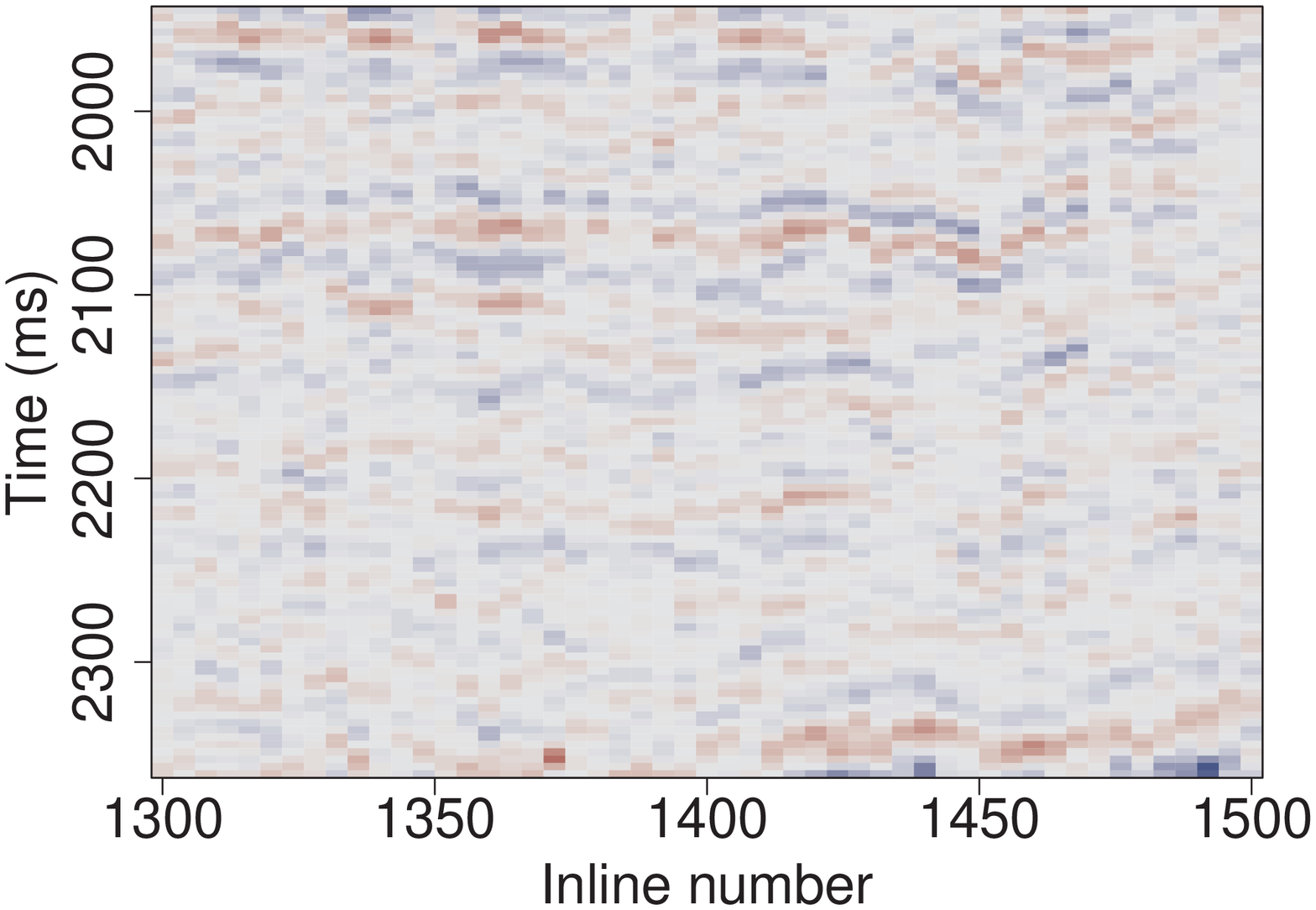}} &
\makebox[6.9cm]{\includegraphics[width=6cm,angle=-90]{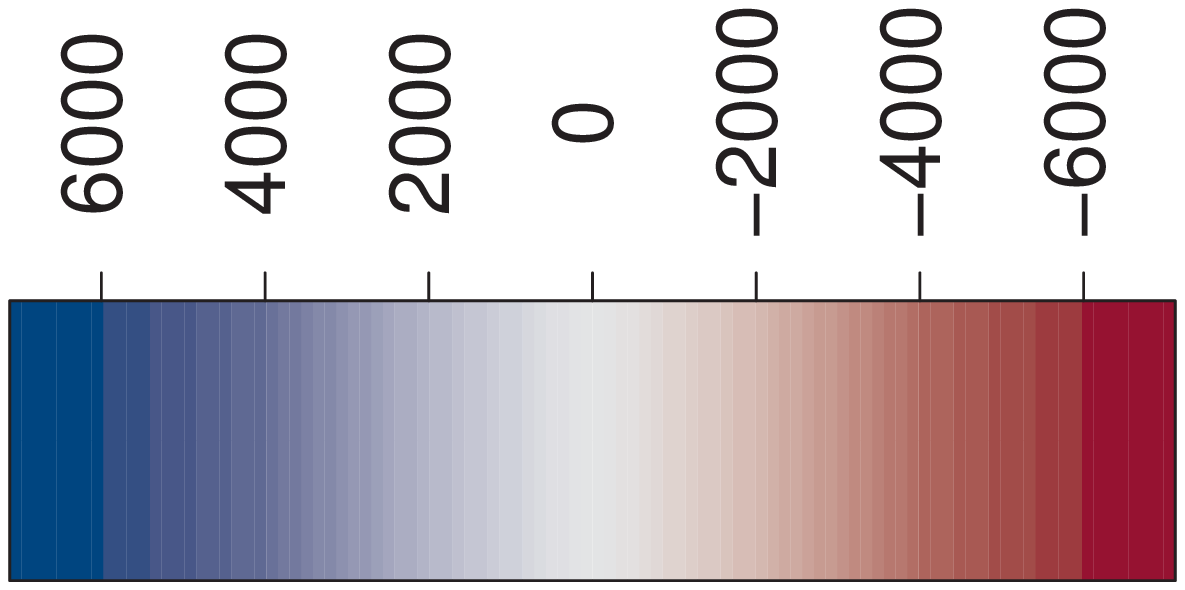}}\\[-0.5cm]
\end{tabular}
\end{center}
\caption{\label{fig:data}Near (left) and far (right) offset seismic data used 
for lithology/fluid prediction.}
\end{figure}
We model two lithology/fluid classes, oil sand and shale. For each node $(i,j)\in G$ we let $\kappa_{ij}
\in \{0,1\}$ denote the lithology fluid class in node $(i,j)$, where $\kappa_{ij}=0$ and $\kappa_{ij}=1$
represent shale and oil sand, respectively. We let $\kappa$ be a vector of all 
$\kappa_{ij},(i,j)\in G$ stacked on top of each other. To estimate a Markov mesh prior distribution 
for $\kappa$ we use a training image from \citet{art173}. The training image is from 
a reservoir with similar characteristics and is shown in Figure \ref{fig:training}. 
\begin{figure}
\begin{center}
\begin{tabular}{c}
\makebox[6.5cm]{\includegraphics[width=5cm,angle=-90]{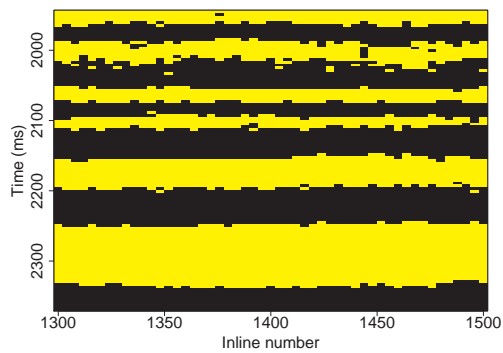}}
\end{tabular}
\end{center}
\caption{\label{fig:training}Training image that we use to estimate a Markov mesh prior distribution 
for the spatial distribution of lithology/fluid classes, $\kappa$. Black and yellow represent 
shale and oil sand, respectively.}
\end{figure}

To model the relation between $\kappa$ and $d$ we first introduce a vector 
$m = \{m_{ij},(i,j)\in G\}$ of elastic properties, where $m_{ij}$ is a vector of length
two. We let the first element in $m_{ij}$ be the product of the density $\rho$ and the pressure-wave velocity 
$v_p$ in node $(i,j)$ and let the second element be the $v_p/v_s$ ratio in the same node, where
$v_s$ is the shear-wave velocity.

For the three variables $\kappa$, $m$ and $d$ we adopt a Bayesian model. We let $p(\kappa)$
denote a prior distribution for $\kappa$ and let $p(m|\kappa)$ denote the conditional 
distribution for the elastic parameters $m$ given the lithology/fluid classes $\kappa$. Finally,
we assume the seismic data $d$ to be conditionally independent of $\kappa$ when 
the elastic properties $m$ are given. We let $p(d|m)$ denote the conditional distribution 
for the seismic data $d$ given elastic properties $m$. The $p(d|m)$ represents a probabilistic
formulation of the forward model. Bayes theorem then gives
\begin{equation}\label{eq:joint}
p(\kappa|d) \propto p(\kappa) p(d|\kappa),
\end{equation}
where 
\begin{equation}
p(d|\kappa) = \int p(m,d|\kappa) \mbox{d}m = \int p(m|\kappa) p(d|m)\mbox{d}m.
\end{equation}
In the following we first outline the details of $p(m|\kappa)$ and $p(d|m)$, 
which is used to specify the likelihood $p(d|\kappa)$, and 
thereafter specify the prior $p(\kappa)$ before we 
describe the Markov chain Monte Carlo procedure we
use to simulate from $p(\kappa|d)$.

\subsection{Likelihood model}
\label{sec:likelihood}
Following \citet{art164} and \citet{tech34} we adopt a linearised and 
Gaussian likelihood for the forward model for $d$ given $\kappa$. More specifically, we assume each of 
$p(m|\kappa)$ and $p(d|m)$ to be Gaussian, the conditional mean of $d$ given $m$ to be a
linear function of $m$, and the conditional covariance matrix of $d$ given $m$ not to 
be a function of $m$. In the following we outline the distributions $p(m|\kappa)$
and $p(d|m)$ in more detail, starting with $p(m|\kappa)$.

We assume $m|\kappa$ to be Gaussian 
and
\begin{equation}
\mbox{E}[m_{ij}|\kappa] = \mu_{\kappa_{ij}} \mbox{~~~~and~~~~}
\mbox{Cov}[m_{ij}|\kappa] = \Sigma_{\kappa_{ij}},
\end{equation}
where $\mu_0$ and $\Sigma_0$ are the conditional mean and covariance for $m_{ij}$ if
node $(i,j)$ contains shale ($\kappa_{ij}=0$), and $\mu_1$ and $\Sigma_1$ are 
corresponding quantities when node $(i,j)$ contains oil sand ($\kappa_{ij}=1$).
Moreover, we assume a separable correlation function $\rho ((i,j),(k,l))$ for 
$m|\kappa$, but we do not allow the correlations to depend on the lithology/fluid classes
$\kappa$. 

For the forward model $p(d|m)$ we use a convolved linearised
approximation of the Zoeppritz equation \citep{art162} based on Aki-Richards formulation that is valid
for weak vertical contrasts \citep{book45}. The vector $d$ is then formed from $m$ 
in several steps. First, all vertical first-order contrasts or 
differences $m_{ij}-m_{i-1,j}$ are formed by pre-multiplying $m$ with a matrix $D$. Thereafter,
reflection coefficients are formed by pre-multiplying $Dm$ with 
a block diagonal matrix $A$, where all the blocks are identical $2\times 2$ matrices
containing coefficients in the Aki-Richards formulation. The mean value of the 
seismic data are then formed via a convolution of each column of $ADm$ with 
wavelets. Different wavelets are used for the near and far offset seismic data as shown in 
Figure \ref{fig:wavelets}. 
\begin{figure}
\begin{center}
\begin{tabular}{ccc}
\makebox[6.5cm]{\includegraphics[width=5cm,angle=-90]{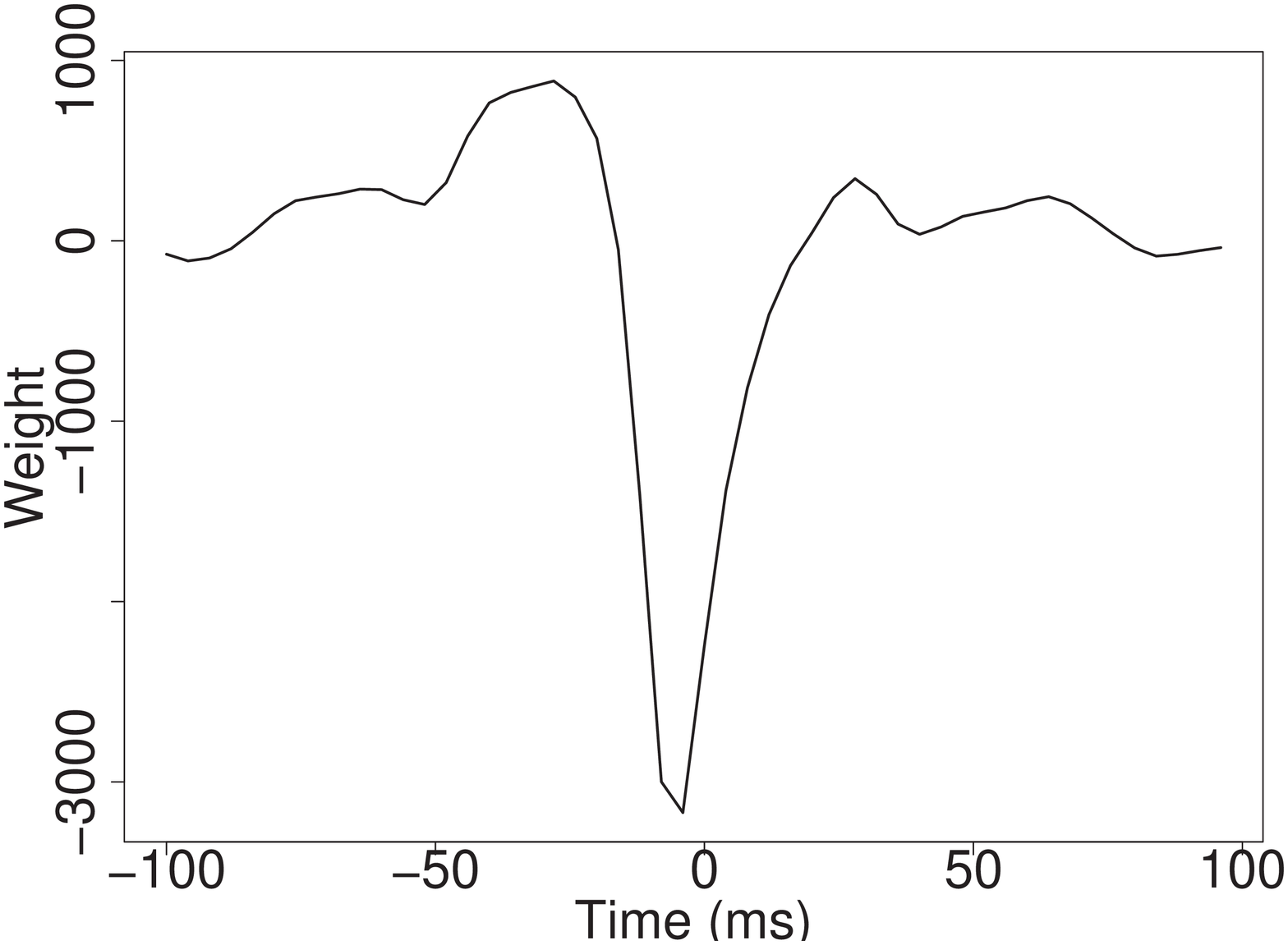}} &
~~~ &
\makebox[6.5cm]{\includegraphics[width=5cm,angle=-90]{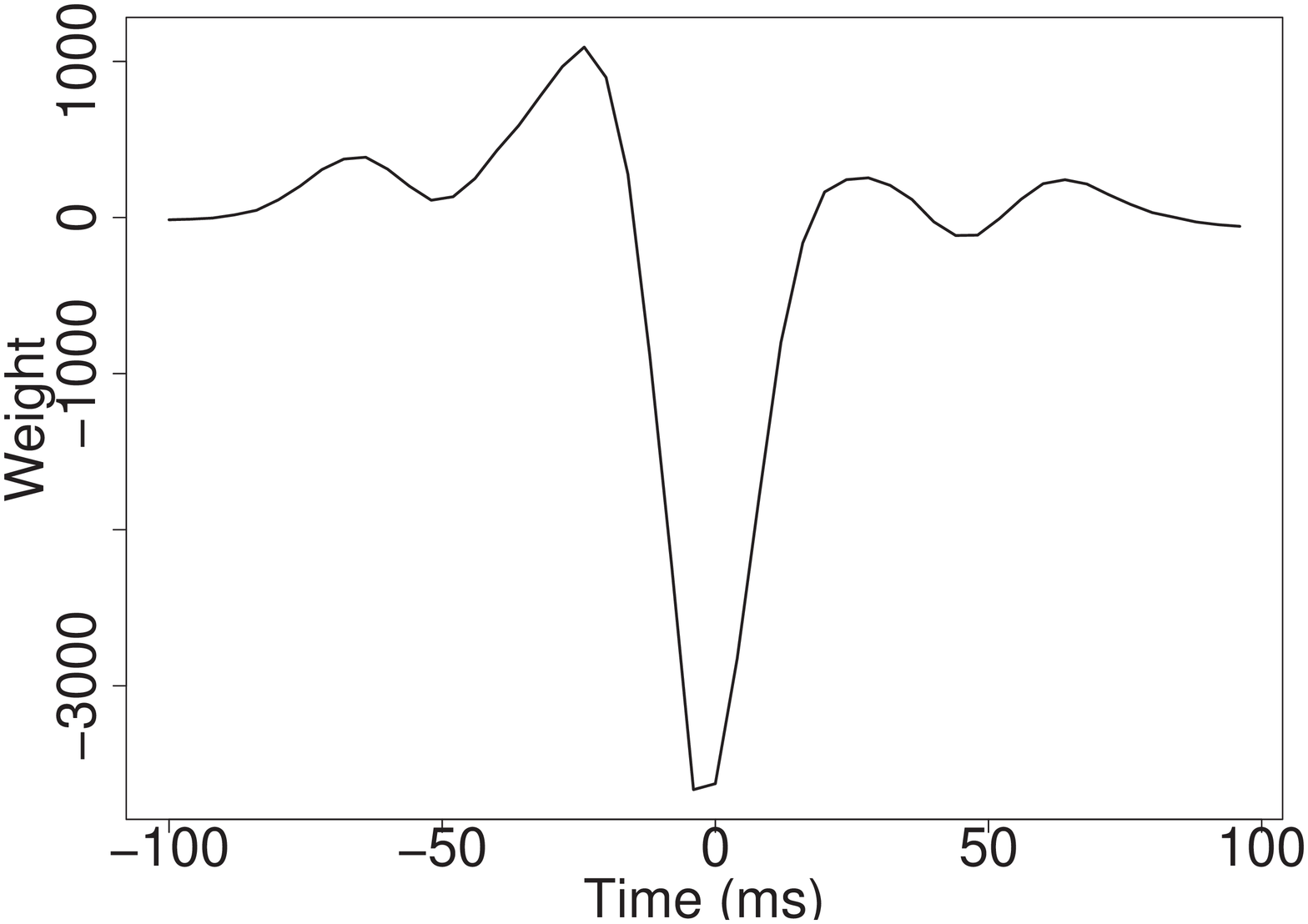}}
\end{tabular}
\end{center}
\caption{\label{fig:wavelets}Wavelets used for the near (left) and far (right) offset seismic data.
The $x$- and $y$- axes show vertical distance and wavelet values, respectively.}
\end{figure}
These wavelets are estimated from data in a well in the same reservoir as the seismic data
is coming from. This well is, however, located some distance away from the seismic section 
we are studying.
The effect of the convolutions can be written as pre-multiplying $ADm$ 
with a matrix $W$. Finally, a zero mean Gaussian error term $\varepsilon$ with a
fixed covariance matrix $\Sigma_\varepsilon$ is added to 
$WADm$. Thus, $d|m$ is Gaussian with 
\begin{equation}
\mbox{E}[d|m] = WADm \mbox{~~~~and~~~~} \mbox{Cov}[d|m] = \Sigma_\varepsilon.
\end{equation}

With the Gaussian distributions specified above for $p(m|\kappa)$ and $p(d|m)$ it
follows from standard properties of the Gaussian distribution that also
$p(d|\kappa)$ becomes Gaussian. Moreover, expressions are available for the mean
$\mbox{E}[d|\kappa]$ and the covariance $\mbox{Cov}[d|\kappa]$ as function of
$\kappa$, $\mu_0$, $\mu_1$, $\Sigma_0$, $\Sigma_1$, $\rho(\cdot,\cdot)$, 
$W$, $A$, $D$ and $\Sigma_\varepsilon$.

\subsection{Prior models}
\label{sec:prior}
The main purpose of this article is to demonstrate how a Markov mesh model fitted to a
training image can be used as prior in a Bayesian model for lithology/fluid prediction. However, 
we also want to study how the inversion results change when using such a prior relative 
to what we get by using a simpler manually specified prior. To fit a Markov mesh model to 
a training image involves extra working and computing time, so there is no reason to do 
so unless it results in a significant change in the inversion results. In the following
we first specify the class of Markov mesh models and briefly discuss the procedure we use to 
fit the model to the training image in Figure \ref{fig:training}. Thereafter we describe
a simpler manually specified prior we use for comparison, the profile Markov random field 
introduced in \citet{art123}.

\subsubsection{Markov mesh prior}
\label{sec:MMprior}
An introduction to the class of Markov mesh models can be found in 
\citet{art133} and the more general class of partially ordered 
Markov models is defined in \citet{art119}. In the following description
we limit the attention to binary fields and introduce the necessary notions 
to define homogeneous Markov mesh models defined on a rectangular lattice.

Let $G= \{(i,j)|i=1,\ldots,m; j=1,\ldots,n\}$ be a rectangular lattice, to each node $(i,j)\in G$ of 
which we associate a binary variable $\kappa_{ij}\in\{0,1\}$. We let $\kappa=(\kappa_{ij}:(i,j)\in G)$ denote 
the collection of all these binary variables and use $\kappa_\lambda=(\kappa_{ij}:(i,j)\in \lambda)$ to denote 
the collection of binary variables in a set $\lambda\subseteq G$ of nodes. The Markov mesh model
is based on numbering the nodes in $G$ from $1$ to $nm$ in the lexicographical order. Without 
loss of generality, the distribution of $\kappa$ can then be expressed as
\begin{equation}
p(\kappa) = \prod_{(i,j)\in G} p(\kappa_{ij}|\kappa_{\rho_{ij}}),
\end{equation}
where $\rho_{ij}$ is the set of all nodes coming before node $(i,j)$, i.e.
\begin{equation}
\rho_{ij} = \{ (k,l)\in G: nk+l < ni+j\}.
\end{equation}
The set $\rho_{ij}$ is called the predecessor set of node $(i,j)$. The central assumption
in Markov mesh models is that $p(\kappa_{ij}|\kappa_{\rho_{ij}})$ has a Markov property in that 
\begin{equation}
p(\kappa_{ij}|\kappa_{\rho_{ij}}) = p(\kappa_{ij}|\kappa_{\nu_{ij}}),
\end{equation}
where $\nu_{ij} \subseteq \rho_{ij}$ is called the sequential neighborhood of node $(i,j)$.
Following \citet{tech35}, we assume that all the sequential neighborhoods are generated via a 
translation of a template sequential neighborhood $\tau$. The set $\tau$ can best be thought of as
the sequential neighborhood of node $(0,0)$ in a infinite lattice. More precisely, $\tau$ should 
contain a finite number of elements and 
\begin{equation}
\tau \subset \{(i,j):i\in \mathbb{Z}^-,j\in\mathbb{Z}\} \cup \{(0,j):j\in\mathbb{Z}^-\},
\end{equation}
where $\mathbb{Z}=\{0,\pm 1,\pm 2,\ldots\}$ and $\mathbb{Z}^- = \{-1,-2,\ldots\}$ are the sets
of all integers and all negative integers, respectively. Given the set $\tau$ we assume the sequential 
neighborhood $\nu_{ij}$ to be generated by translating each element in $\tau$ a distance $(i,j)$ and,
if necessary, dropping elements falling outside the lattice $G$. Mathematically, $\nu_{ij}$ is then given 
as 
\begin{equation}
\nu_{ij} = (\tau \oplus (i,j)) \cap G,
\end{equation}
where the translation operator $\oplus$ is defined as
\begin{equation}
\tau \oplus (i,j) = \{(k+i,l+j):(k,l)\in \tau\}. 
\end{equation}
Constructing $\nu_{ij}$ in this way, the sequential neighborhoods for all nodes sufficiently far away from 
the lattice borders will have the same form.

Still following \citet{tech35}, we model $p(\kappa_{ij}|\kappa_{\nu_{ij}})$ by assuming the logit 
transformation of $p(\kappa_{ij}=1|\kappa_{\nu_{ij}})$ to be given by
\begin{equation}
\mbox{logit}\left[ p(\kappa_{ij}=1|\kappa_{\nu_{ij}})\right] = \ln \left( \frac{p(\kappa_{ij}|\kappa_{\nu_{ij}})}{1-p(\kappa_{ij}=1|\kappa_{\nu_{ij}})}\right) = 
\theta (\xi (\kappa,\tau,(i,j))),
\end{equation}
where $\xi(\kappa,\tau,(i,j))\subseteq \tau$ is the set of elements $(k,l)\in \tau$ associated to a node with oil sand
in the sequential neighborhood for node $(i,j)$, i.e.
\begin{equation}
\xi (\kappa,\tau,(i,j)) = \{ (k,l)\in \tau : (i+k,j+l)\in G \mbox{~and~} \kappa_{i+k,j+l}=1\},
\end{equation}
and $\theta(\cdot )$ is a pseudo-Boolean function \citep{art137,art138} 
to be specified. One should note that, as we assume the same 
function $\theta(\cdot )$ for all nodes $(i,j)\in G$ we get a homogeneous model. Moreover, one should note that 
the definition of $\xi(\kappa,\tau,(i,j))$ implies that for nodes $(i,j)$ close to the boundary of the lattice, so that 
$(\tau \oplus (i,j))\setminus G \neq \emptyset$, the conditional distribution $p(\kappa_{ij}|\kappa_{\nu_{ij}})$ becomes as if
one had an infinite lattice where all variables associated to nodes outside of $G$ were zero. 

The last step in specifying the Markov mesh model is to choose the function $\theta(\cdot)$. This is a real 
valued function, where the argument is a subset of $\tau$ specifying for which sequential neighbors the associated 
binary variable is equal to one. Without loss of generality, the $\theta(\cdot )$ can be uniquely 
expressed in terms of a collection of interaction parameters $\{\beta(\lambda): \lambda\subseteq \tau\}$ by
\begin{equation}
\theta(\lambda) = \sum_{\lambda^\star\subseteq \lambda}\beta(\lambda^\star) \mbox{~~for $\lambda\subseteq \tau$.}
\end{equation}
The number of interaction parameters is $2^{|\tau|}$, where $|\tau|$ is the number of elements in $\tau$. Unless
$|\tau|$ is very small the number of parameters necessary to specify $\theta(\cdot )$ is thereby very large.
We still follow \citet{tech35} and limit the number of model parameters by restricting many of the interaction 
parameters to be zero. More specifically,
for some $\Lambda\subseteq\Omega(\tau)$, where $\Omega(\tau)$ is the set of all subsets of $\tau$, 
we assume $\beta(\lambda)=0$ for all $\lambda\not\in \Lambda$. Thus, we specify the Markov mesh model by 
choosing the sets $\tau$ and $\Lambda$ and the interaction values $\{\beta(\lambda):\lambda\in\Lambda\}$.

To fit the Markov mesh model specified above to the training image in Figure \ref{fig:training} we adopt 
the Bayesian procedure introduced in \citet{tech35}, including the hyper-parameter values used in that article.
A prior is specified for $\tau$, $\Lambda$ and $\{\beta(\lambda):\lambda\in\Lambda\}$ and 
assuming the training image to be a sample from the specified Markov mesh model,
a Metropolis--Hastings algorithm is used to generate samples from the resulting posterior distribution 
for $\tau$, $\Lambda$ and $\{\beta(\lambda):\lambda\in\Lambda\}$. 
When we conditioned on the training image in Figure \ref{fig:training},
the convergence of the Metropolis--Hastings algorithm was so slow that we were unable
to obtain convergence within a reasonable computation time. As a pragmatic approach to obtain a reasonable 
prior $p(\kappa)$ we simply run the Metropolis--Hastings
algorithm in \citet{tech35} for a large number of iterations and used the last values for $\tau$, $\Delta$ and 
$\{\beta(\lambda):\lambda\in\Lambda\}$ in this run to define the prior $p(\kappa)$. The resulting $p(\kappa)$ prior
has $|\tau|=9$ sequential neighbors and $|\Delta| = 31$ interaction parameters that are allowed to differ 
from zero. The sequential neighborhood $\tau$ is illustrated in Figure \ref{fig:tau}, while the 
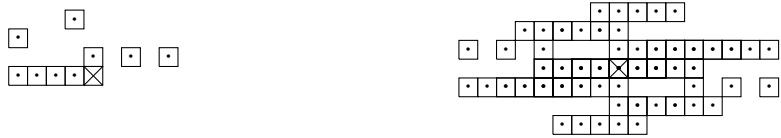
\begin{figure}
\begin{center}
\begin{tabular}{ccc}
\begin{tikzpicture}[scale=0.25]
\draw[color=black] (0,0) -- (0,1) -- (1,1) -- (1,0) -- (0,0);
\draw[color=black] (0,0) -- (1,1);
\draw[color=black] (0,1) -- (1,0);

\draw[color=black] (0,1) -- (1,1) -- (1,2) -- (0,2) -- (0,1);
\draw[color=black] (-1,0) -- (0,0) -- (0,1) -- (-1,1) -- (-1,0);
\draw[color=black] (2,1) -- (3,1) -- (3,2) -- (2,2) -- (2,1);
\draw[color=black] (-2,0) -- (-1,0) -- (-1,1) -- (-2,1) -- (-2,0);
\draw[color=black] (-1,3) -- (0,3) -- (0,4) -- (-1,4) -- (-1,3);
\draw[color=black] (-3,0) -- (-2,0) -- (-2,1) -- (-3,1) -- (-3,0);
\draw[color=black] (4,1) -- (5,1) -- (5,2) -- (4,2) -- (4,1);
\draw[color=black] (-4,0) -- (-3,0) -- (-3,1) -- (-4,1) -- (-4,0);
\draw[color=black] (-4,2) -- (-3,2) -- (-3,3) -- (-4,3) -- (-4,2);

\draw[color=black] (0+0.5,1+0.5) node {$\cdot$}; 
\draw[color=black] (-1+0.5,0+0.5) node {$\cdot$}; 
\draw[color=black] (2+0.5,1+0.5) node {$\cdot$}; 
\draw[color=black] (-2+0.5,0+0.5) node {$\cdot$}; 
\draw[color=black] (-1+0.5,3+0.5) node {$\cdot$}; 
\draw[color=black] (-3+0.5,0+0.5) node {$\cdot$}; 
\draw[color=black] (4+0.5,1+0.5) node {$\cdot$}; 
\draw[color=black] (-4+0.5,0+0.5) node {$\cdot$}; 
\draw[color=black] (-4+0.5,2+0.5) node {$\cdot$}; 

\draw[color=white] (0,-3) -- (1,-3);
\end{tikzpicture}
&\mbox{~~~~~~~~~~~~~~~~~~~~} &
\begin{tikzpicture}[scale=0.25]
\draw[color=black] (0,0) -- (0,1) -- (1,1) -- (1,0) -- (0,0);
\draw[color=black] (0,0) -- (1,1);
\draw[color=black] (0,1) -- (1,0);

\foreach \x/\y in {0/0,-1/0,0/-1,-1/2,0/-2,-3/-1,0/-3,-1/4,0/-4,-2/-4} {
  \draw[color=black] (0-\y,0+\x) -- (0-\y,1+\x) -- (1-\y,1+\x) -- (1-\y,0+\x) -- (0-\y,0+\x);
  \draw[color=black] (0-\y,1+\x) -- (1-\y,1+\x) -- (1-\y,2+\x) -- (0-\y,2+\x) -- (0-\y,1+\x);
  \draw[color=black] (-1-\y,0+\x) -- (0-\y,0+\x) -- (0-\y,1+\x) -- (-1-\y,1+\x) -- (-1-\y,0+\x);
  \draw[color=black] (2-\y,1+\x) -- (3-\y,1+\x) -- (3-\y,2+\x) -- (2-\y,2+\x) -- (2-\y,1+\x);
  \draw[color=black] (-2-\y,0+\x) -- (-1-\y,0+\x) -- (-1-\y,1+\x) -- (-2-\y,1+\x) -- (-2-\y,0+\x);
  \draw[color=black] (-1-\y,3+\x) -- (0-\y,3+\x) -- (0-\y,4+\x) -- (-1-\y,4+\x) -- (-1-\y,3+\x);
  \draw[color=black] (-3-\y,0+\x) -- (-2-\y,0+\x) -- (-2-\y,1+\x) -- (-3-\y,1+\x) -- (-3-\y,0+\x);
  \draw[color=black] (4-\y,1+\x) -- (5-\y,1+\x) -- (5-\y,2+\x) -- (4-\y,2+\x) -- (4-\y,1+\x);
  \draw[color=black] (-4-\y,0+\x) -- (-3-\y,0+\x) -- (-3-\y,1+\x) -- (-4-\y,1+\x) -- (-4-\y,0+\x);
  \draw[color=black] (-4-\y,2+\x) -- (-3-\y,2+\x) -- (-3-\y,3+\x) -- (-4-\y,3+\x) -- (-4-\y,2+\x);

  \draw[color=black] (0-\y+0.5,0+\x+0.5) node {$\cdot$}; 
  \draw[color=black] (0-\y+0.5,1+\x+0.5) node {$\cdot$}; 
  \draw[color=black] (-1-\y+0.5,0+\x+0.5) node {$\cdot$};
  \draw[color=black] (2-\y+0.5,1+\x+0.5) node {$\cdot$}; 
  \draw[color=black] (-2-\y+0.5,0+\x+0.5) node {$\cdot$};
  \draw[color=black] (-1-\y+0.5,3+\x+0.5) node {$\cdot$};
  \draw[color=black] (-3-\y+0.5,0+\x+0.5) node {$\cdot$};
  \draw[color=black] (4-\y+0.5,1+\x+0.5) node {$\cdot$}; 
  \draw[color=black] (-4-\y+0.5,0+\x+0.5) node {$\cdot$};
  \draw[color=black] (-4-\y+0.5,2+\x+0.5) node {$\cdot$};
}
\end{tikzpicture}
\end{tabular}
\end{center}
\caption{\label{fig:tau}The sequential neighborhood (left) and the corresponding Markov random field neighborhood
(right) for the fitted prior $p(\kappa)$. The nodes marked with a dot are (sequential) neighbors of the node
marked with a cross.}
\end{figure}
complete specification of $\tau$, $\Delta$ and $\{\beta(\lambda):\lambda\in\Lambda\}$ is given in 
Appendix \ref{app:prior}. The best way to understand the properties of the prior is perhaps to look at 
realisations sampled from $p(\kappa)$, four of which are shown in Figure \ref{fig:realPrior}.
\begin{figure}
\begin{center}
\begin{tabular}{@{}cc@{}}
\makebox[6.5cm]{\includegraphics[width=5cm,angle=-90]{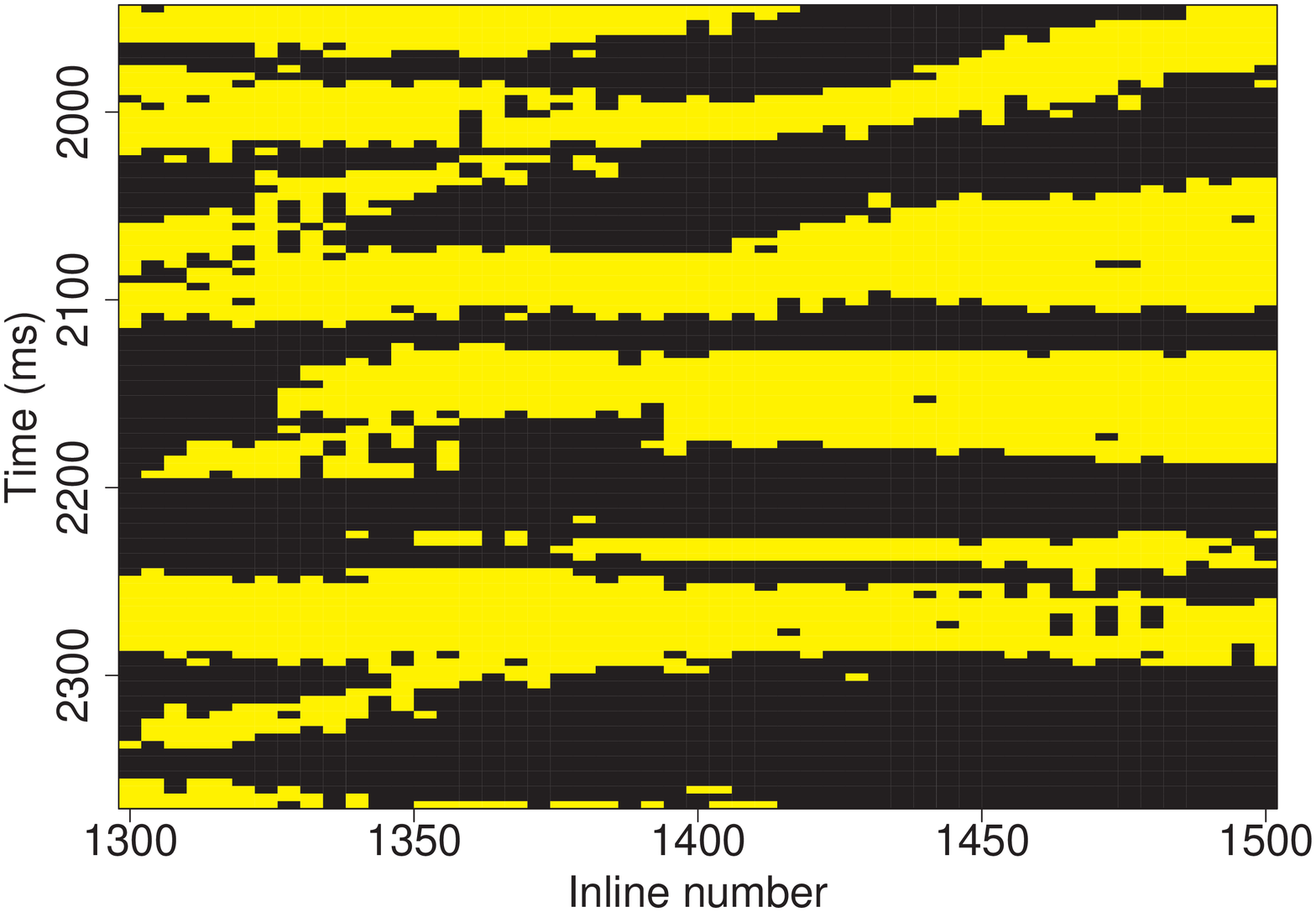}} &
\makebox[6.5cm]{\includegraphics[width=5cm,angle=-90]{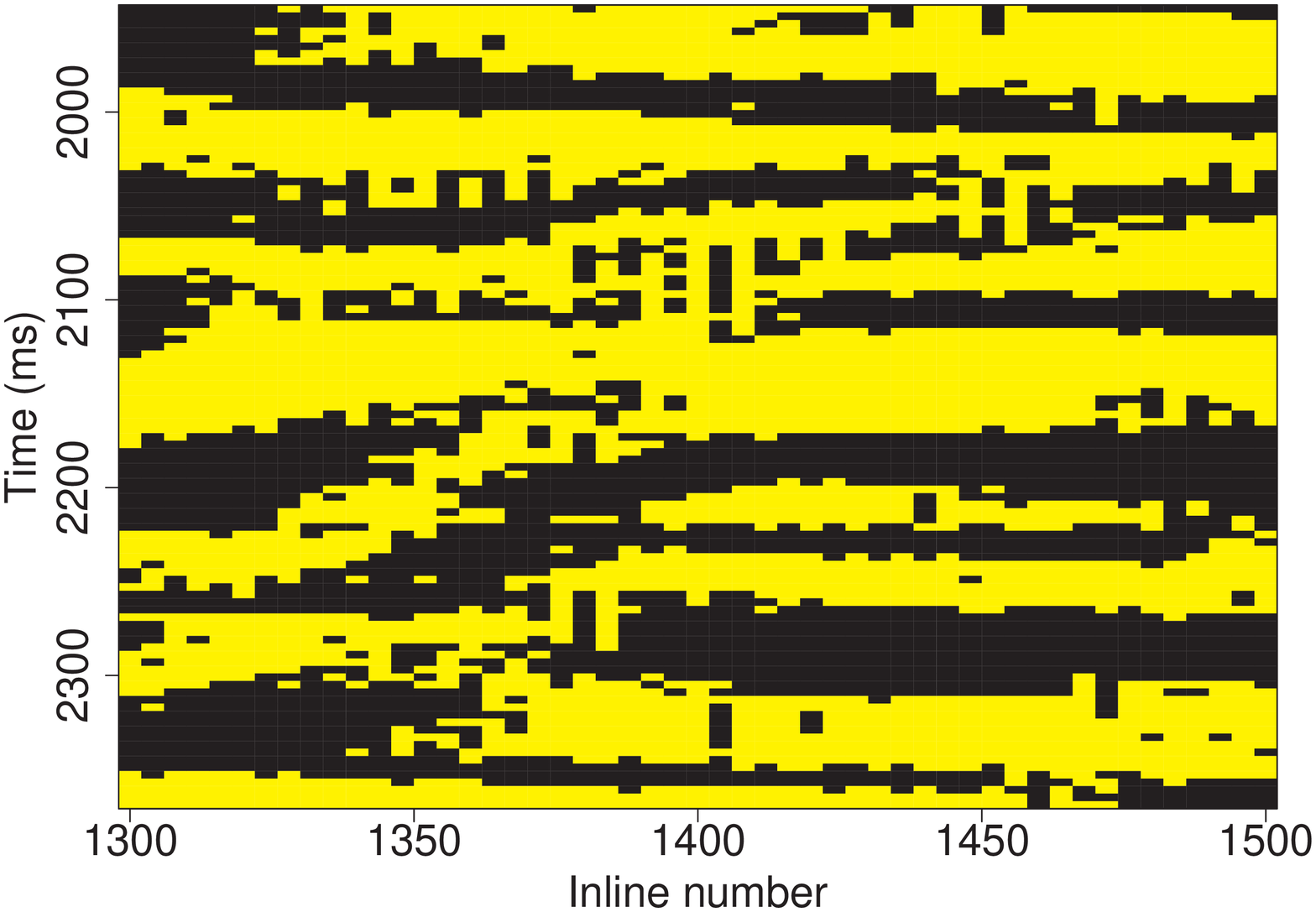}} \\[-0.3cm]
\makebox[6.5cm]{\includegraphics[width=5cm,angle=-90]{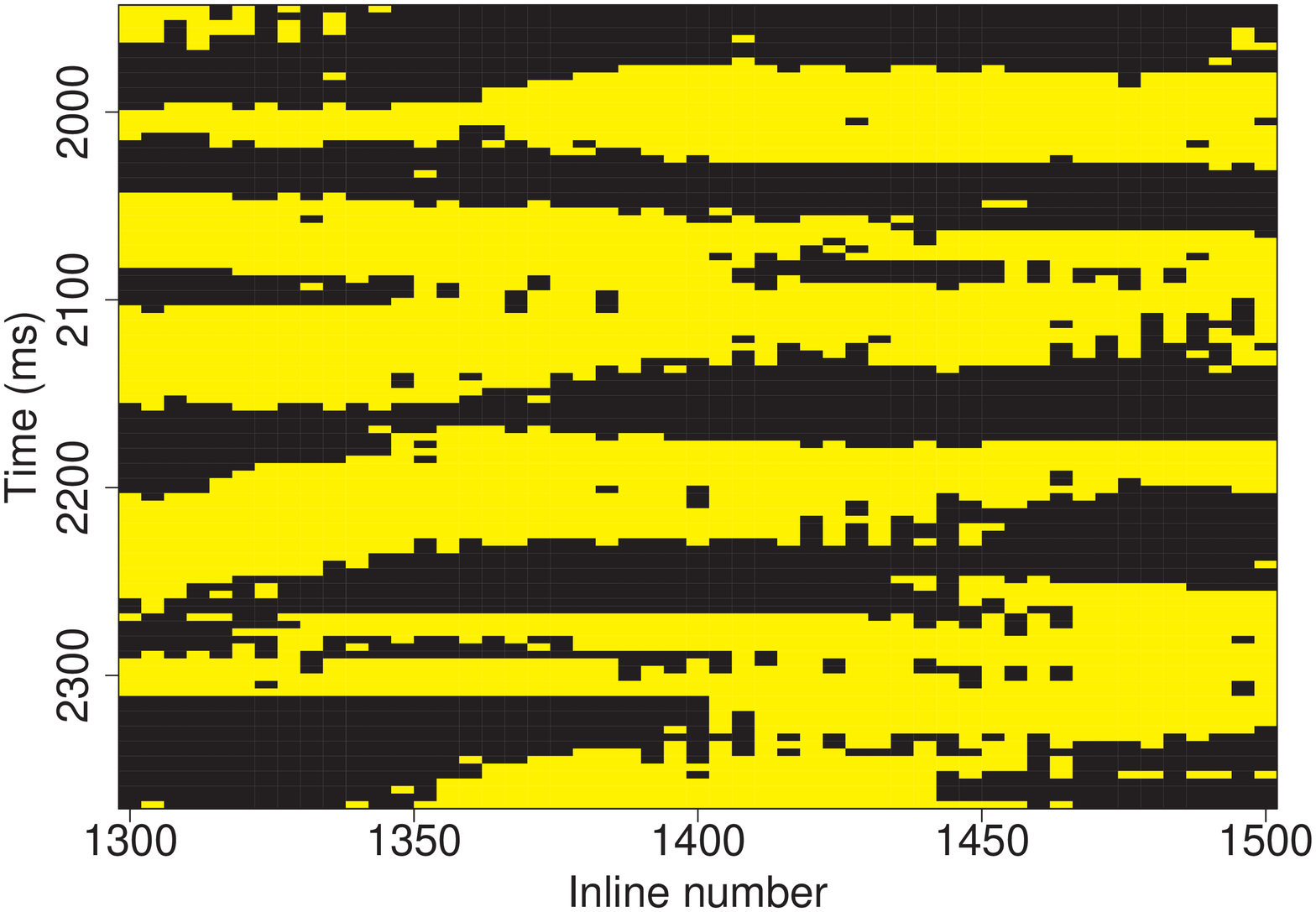}} &
\makebox[6.5cm]{\includegraphics[width=5cm,angle=-90]{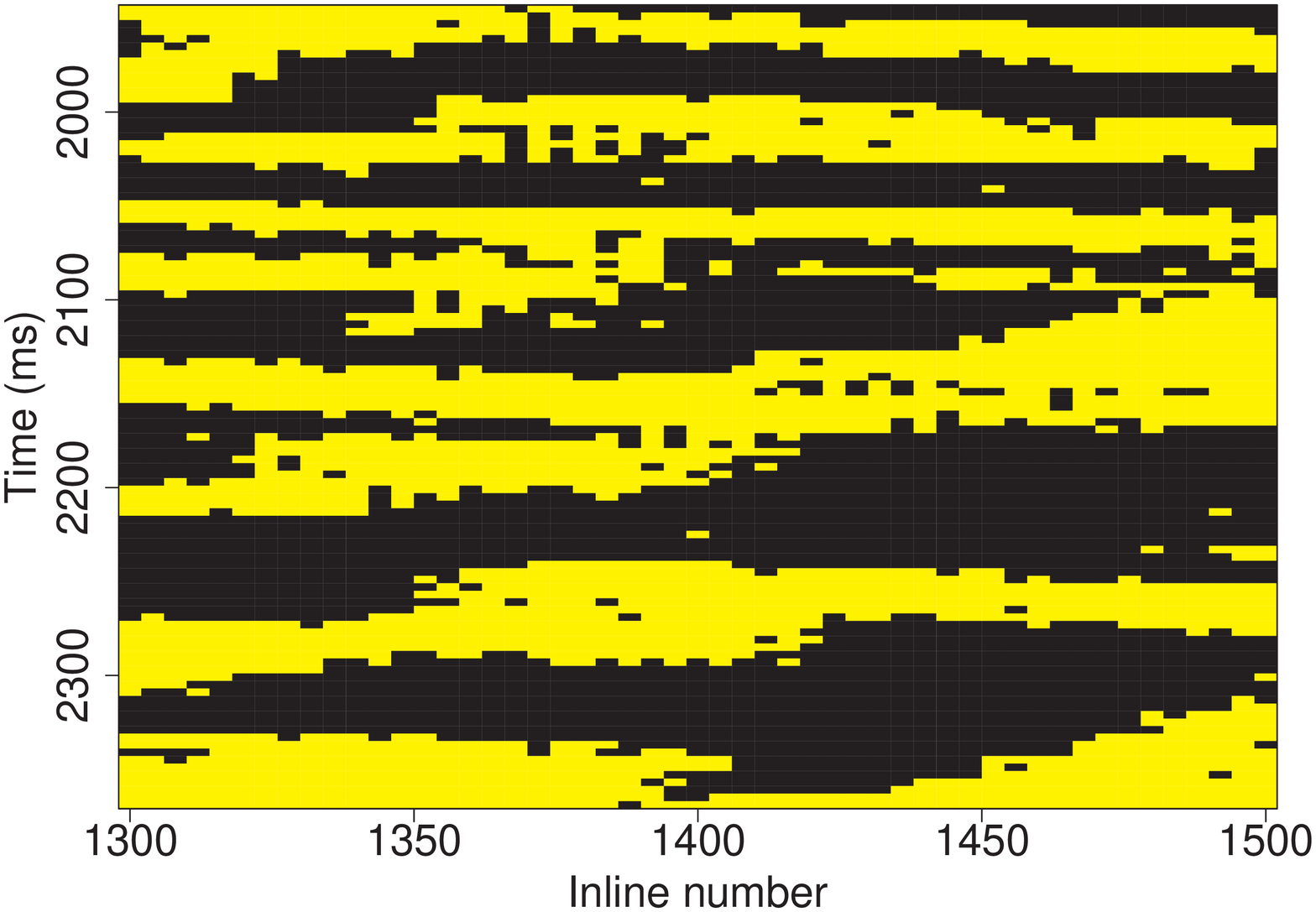}}
\end{tabular}
\end{center}
\caption{\label{fig:realPrior}Four independent realisations from the Markov mesh prior fitted to 
the training image shown in Figure \ref{fig:training}. Black and yellow represent 
shale and oil sand, respectively.}
\end{figure}
We see that the fitted prior is reproducing large continuous areas of shale and oil sand as seen in the 
training image, but the boundaries between shale and oil sand is less horizontal in the realisations from 
the prior than in the training image.

\subsubsection{Profile Markov random field prior}
\label{sec:profile}
The profile Markov random field prior was first defined and used for seismic inversion in 
\citet{art123}, see also \citet{art171}. Even though the prior class is defined for 
categorical variables, in our description of the model we limit the attention to 
the binary variable case.

Let again $G= \{(i,j)|i=1,\ldots,m; j=1,\ldots,n\}$ be a rectangular lattice,
where in each node we associate a binary variable $\kappa_{ij}\in\{0,1\}$. We let $C_j=\{(i,j):i=1,\ldots,m\}$
be the set of nodes in profile or column $j$ of the lattice $G$ and let $\kappa_{C_j} = (\kappa_{ij}:(i,j)\in C_j)$ 
denote the collection of the binary variables associated to this column. The collection of all the binary 
variables except the ones in column $j$ we denote by $\kappa_{-C_j}$. The profile Markov random field
prior is then specified by first adopting the Markov property
\begin{equation}\label{eq:profile}
p(\kappa_{C_j}|\kappa_{-C_j}) = p(\kappa_{C_j}|\kappa_{C_{j-1}},\kappa_{C_{j+1}}),
\end{equation}
i.e. given the values in columns $j-1$ and $j+1$, the values in column $j$ are independent of the 
values in the remaining columns. Secondly, the profile Markov random field prior assumes
$p(\kappa_{C_j}|\kappa_{C_{j-1}},\kappa_{C_{j+1}})$ to be a Markov chain down along the column,
\begin{equation}
p(\kappa_{C_j}|\kappa_{C_{j-1}},\kappa_{C_{j+1}}) = p(\kappa_{(1,j)}|\kappa_{(1,j-1)},\kappa_{(1,j+1)})
 \times \prod_{i=2}^n p(\kappa_{(i,j)}|\kappa_{(i-1,j)},\kappa_{(i,j-1)},\kappa_{(i,j+1)}),
\end{equation}
where the conditional distribution $p(\kappa_{(i,j)}|\kappa_{(i-1,j)},\kappa_{(i,j-1)},\kappa_{(i,j+1)})$ is 
the same for all values of $i$ and $j$. The values we have used for the transition probabilities 
are defined in Table \ref{tab:profile}.
\begin{table}
\caption{\label{tab:profile}Values used for $p(\kappa_{(i,j)}|\kappa_{(i-1,j)},\kappa_{(i,j-1)},\kappa_{(i,j+1)})$ in the 
specification of the profile Markov random field prior.}
\begin{center}
\begin{tabular}{ccc}
$\kappa_{i,j-1}=0,\kappa_{i,j+1}=0$ & ~~~~~~~ & $\kappa_{i,j-1}=0,\kappa_{i,j+1}=1$ \\[0.2cm]
\begin{tabular}{c|cc}
 & $\kappa_{ij}=0$ & $\kappa_{ij}=1$ \\ \hline
$\kappa_{i-1,j}=0$ & 0.9877 & 0.0123 \\
$\kappa_{i-1,j}=1$ & 0.8339 & 0.1661 \\
\end{tabular}
& &
\begin{tabular}{c|cc}
 & $\kappa_{ij}=0$ & $\kappa_{ij}=1$ \\ \hline
$\kappa_{i-1,j}=0$ & 0.6539 & 0.3461 \\
$\kappa_{i-1,j}=1$ & 0.1056 & 0.8944 \\
\end{tabular} \\
~\\
$\kappa_{i,j-1}=1,\kappa_{i,j+1}=0$ & ~~~~~~~ & $\kappa_{i,j-1}=1,\kappa_{i,j+1}=1$ \\[0.2cm]
\begin{tabular}{c|cc}
 & $\kappa_{ij}=0$ & $\kappa_{ij}=1$ \\ \hline
$\kappa_{i-1,j}=0$ & 0.6539 & 0.3461 \\
$\kappa_{i-1,j}=1$ & 0.1056 & 0.8944 \\
\end{tabular}
& &
\begin{tabular}{c|cc}
 & $\kappa_{ij}=0$ & $\kappa_{ij}=1$ \\ \hline
$\kappa_{i-1,j}=0$ & 0.0425 & 0.9575 \\
$\kappa_{i-1,j}=1$ & 0.0028 & 0.9972 \\
\end{tabular}
\end{tabular}
\end{center}
\end{table}
\citet{art171} describe the structure used to specify these values. The basic idea is 
that these values  should represent high probability for lateral continuity of oil sand and
shale. The initial distribution $p(\kappa_{(1,j)}|\kappa_{(1,j-1)},\kappa_{(1,j+1)})$ are set equal to 
$p(\kappa_{(i,j)}|\kappa_{(i-1,j)}=0,\kappa_{(i,j-1)},\kappa_{(i,j+1)})$, i.e. conditioning on shale being 
present above the lattice. Correspondingly, $p(\kappa_{(i,j)}|\kappa_{(i-1,j)},\kappa_{(i,j-1)},\kappa_{(i,j+1)})$
for the left and rightmost columns $j=1$ and $j=n$ are defined by conditioning on shale being present outside
the lattice. 

\subsection{Posterior model and simulation algorithm}
\label{sec:posterior}

For each of the Markov mesh and profile Markov random field priors we obtain a 
posterior distribution for the lithology/fluid classes $\kappa$, given in (\ref{eq:joint}).
To explore and estimate properties of the two posterior distributions we adopt the Metropolis--Hastings 
algorithm \citep{book17,book24,book30}. Since the wavelets in the likelihood model induce strong
dependencies between different $\kappa_{ij}$'s in the same column, a simple single-site updating scheme would
give a Markov chain with a long burn-in and slow mixing. We therefore instead adopt the 
proposal scheme previously used in \citet{art171} and propose in each iteration new values for 
all lithology/fluid classes in one column. Using notation from the discussion of the profile 
Markov random field prior, the joint full conditional for the lithology/fluid classes in column $j$ is
\begin{equation}\label{eq:fullcond}
p(\kappa_{C_j}|\kappa_{-C_j},d) \propto p(\kappa_{C_j}|\kappa_{-C_j}) p(d|\kappa).
\end{equation}
To sample from this distribution is, however, computationally very expensive due to the long range 
dependencies in $\kappa_{C_j}$ induced by the wavelets in the likelihood model. Still following
\citet{art171} we therefore adopt the approximation scheme specified in \citet{art169} to 
construct an approximation $p^\star_\nu(\kappa_{C_j}|\kappa_{-C_j},d)$ to
$p(\kappa_{C_j}|\kappa_{-C_j},d)$, where $\nu$ is an algorithmic tuning parameter, and 
generate potential new values for $\kappa_{C_j}$ by sampling from $p^\star_\nu(\kappa_{C_j}|\kappa_{-C_j},d)$. The 
$p^\star_\nu(\kappa_{C_j}|\kappa_{-C_j},d)$ is a higher-order Markov chain and thereby by 
construction easy to sample from. In general the approximation quality grows with $\nu$, 
but so does the computation 
time required for simulating one realisation from $p^\star_\nu(\kappa_{C_j}|\kappa_{-C_j},d)$. Based on preliminary 
runs of the Metropolis--Hastings algorithm we find a value for $\nu$ which gives reasonable 
acceptance rates for the Metropolis--Hastings algorithm.

To run the Metropolis--Hastings scheme discussed above we first need to have expressions for 
$p(\kappa_{C_j}|\kappa_{-C_j})$ for each of the two priors. For the profile Markov random field prior
this is by construction given by (\ref{eq:profile}) and the values in Table \ref{tab:profile}.
To obtain $p(\kappa_{C_j}|\kappa_{-C_j})$ for the Markov mesh prior we first need to reformulate
the Markov mesh model as a Markov random field. The resulting 
Markov random field has a neighborhood system where the set of neighbors to node $(i,j)\in G$ is
\begin{equation}
\partial_{ij} = \nu_{ij} \cup \left( \bigcup_{(k,l)\in G:(i,j)\in\nu_{kl}} \left(\nu_{kl}\cup \{(k,l)\}\right)\right).
\end{equation}
For nodes sufficiently far away from the lattice borders the $\partial_{ij}$ becomes as shown in Figure \ref{fig:tau}(b).
When the prior $p(x)$ is formulated as a Markov random field it is straightforward to find the corresponding 
$p(\kappa_{C_j}|\kappa_{-C_j})$ by first ignoring potential functions for cliques which do not include any node in $C_j$ 
and thereafter plugging in values for $x_{-C_j}$ in the remaining potential functions. 
In particular, for the Markov mesh sequential neighborhood used 
in this study, $p(\kappa_{C_j}|\kappa_{-C_j})$ becomes a third-order Markov chain. 

The second factor in (\ref{eq:fullcond}) is a high dimensional multivariate Gaussian density. To be able to evaluate
this efficiently it is essential that we have chosen the correlation structure of $d|\kappa$ to be separable.
For each of the two priors we run the Metropolis--Hastings algorithm scheme for the resulting posterior
distribution for a large number of iterations. We use standard output analysis to identify and discard a burn-in 
period. In the next section we use the $\kappa$ realisations after the burn-in period to estimate and compare properties of 
the two posterior distributions.

\section{North Sea Case study}

Recall that the objective is to assess the posterior of the lithology/fluid classes $\kappa_{ij} \in G$ given seicmic AVO data $d$ in a clastic oil reservoir in the North Sea. That is, we want to assess the posterior $p(\kappa|d)$ given in Equation~\ref{eq:joint}, for the two prior models discussed earlier. Note that the two posteriors will not be identical since the priors are different.

\label{sec:results}
To study and compare the properties of the two posterior distributions we can first look at the posterior realisations shown 
in Figure \ref{fig:realPosterior}.
\begin{figure}
\begin{center}
\begin{tabular}{@{}cc@{}}\\[-2.0cm]
\makebox[6.5cm]{\includegraphics[width=5cm,angle=-90]{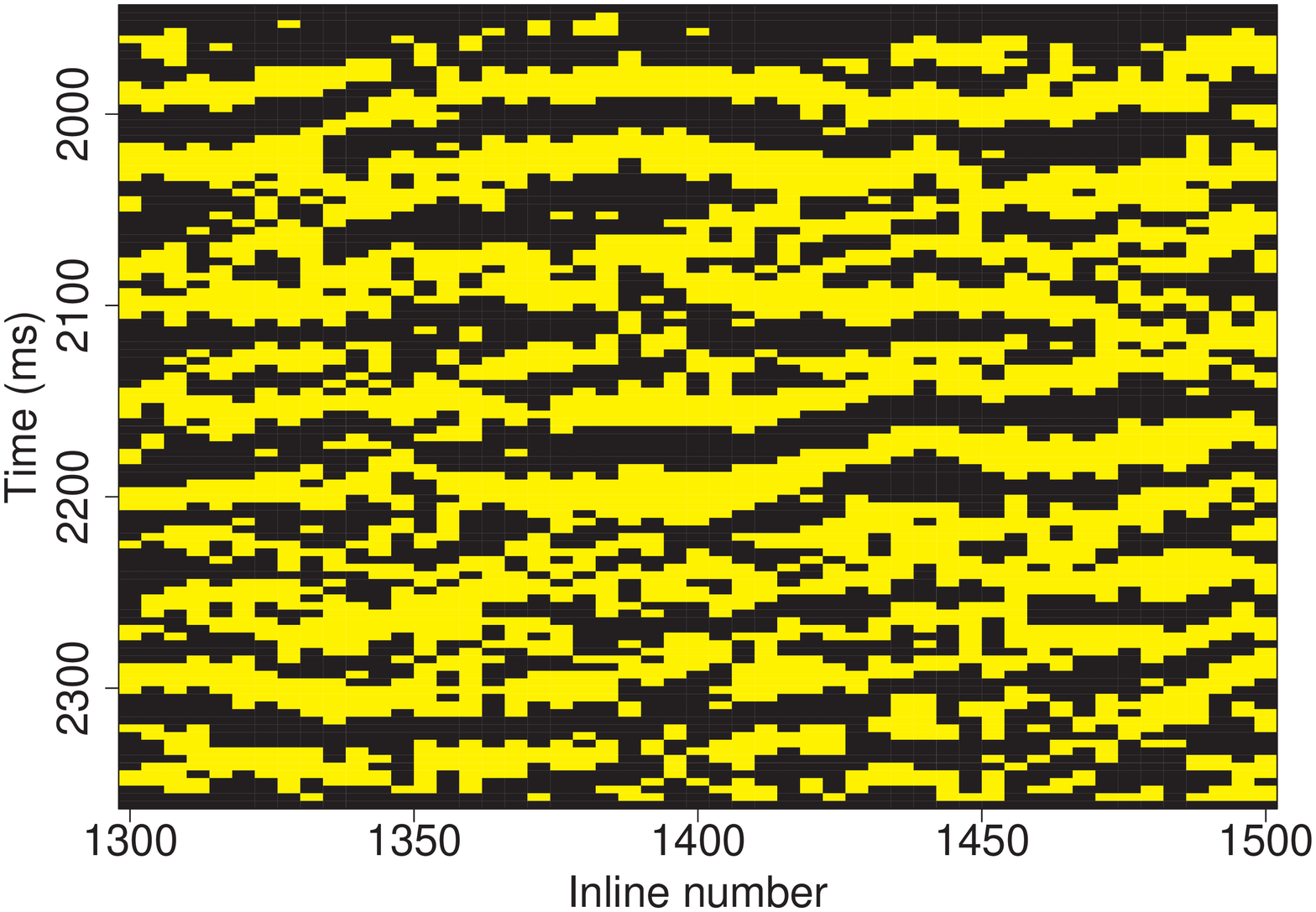}} &
\makebox[6.5cm]{\includegraphics[width=5cm,angle=-90]{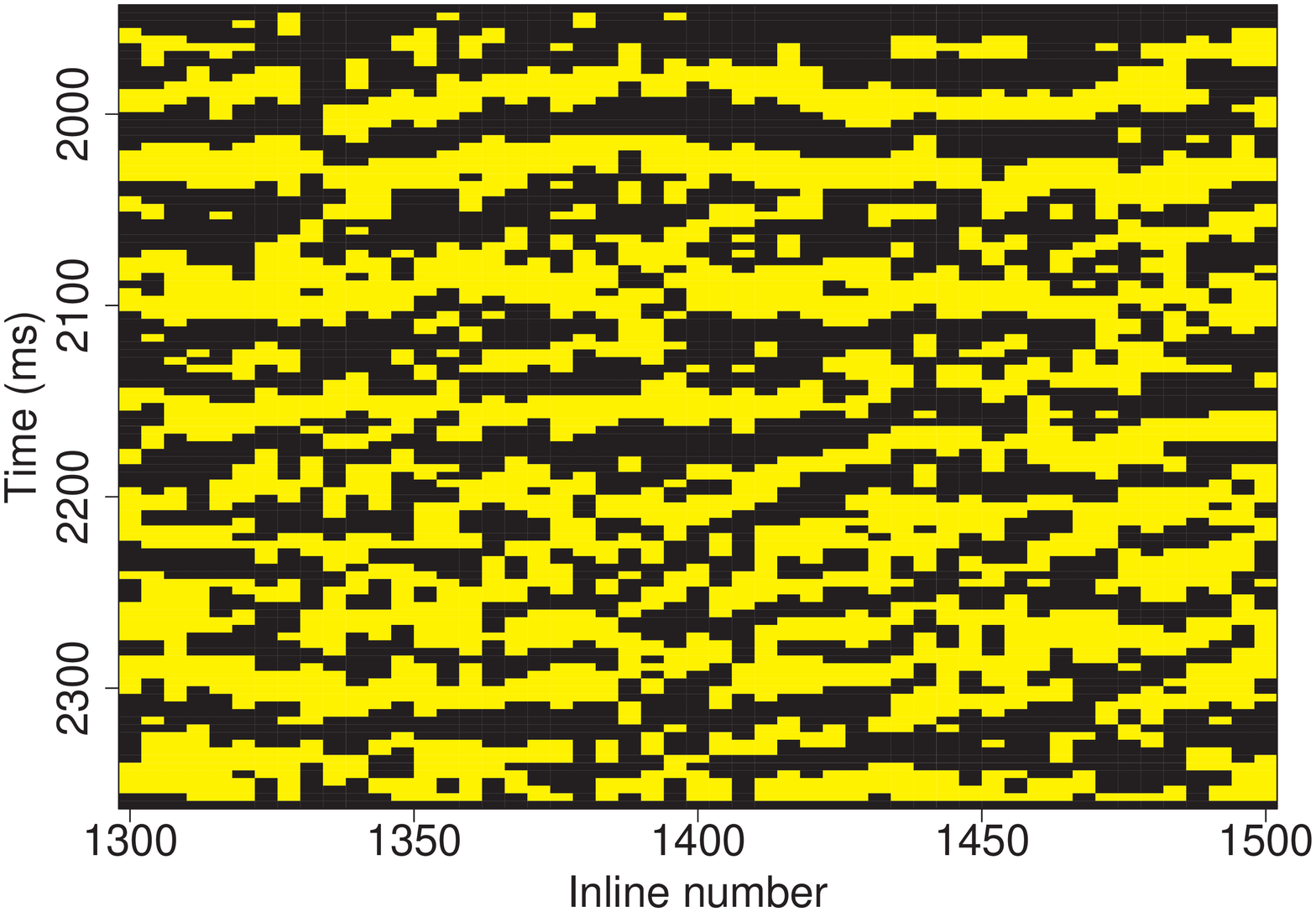}} \\[-0.3cm]
\makebox[6.5cm]{\includegraphics[width=5cm,angle=-90]{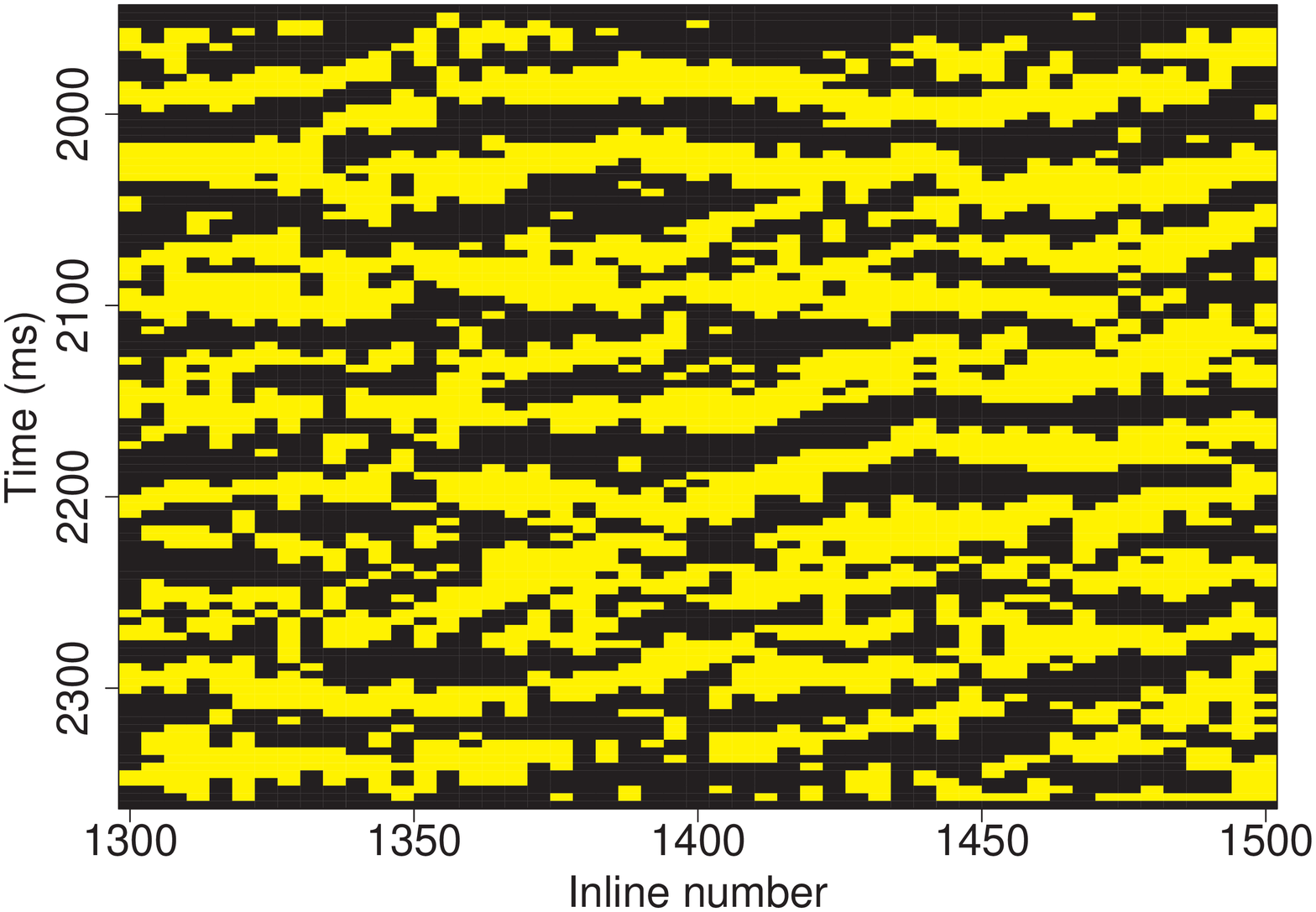}} &
\makebox[6.5cm]{\includegraphics[width=5cm,angle=-90]{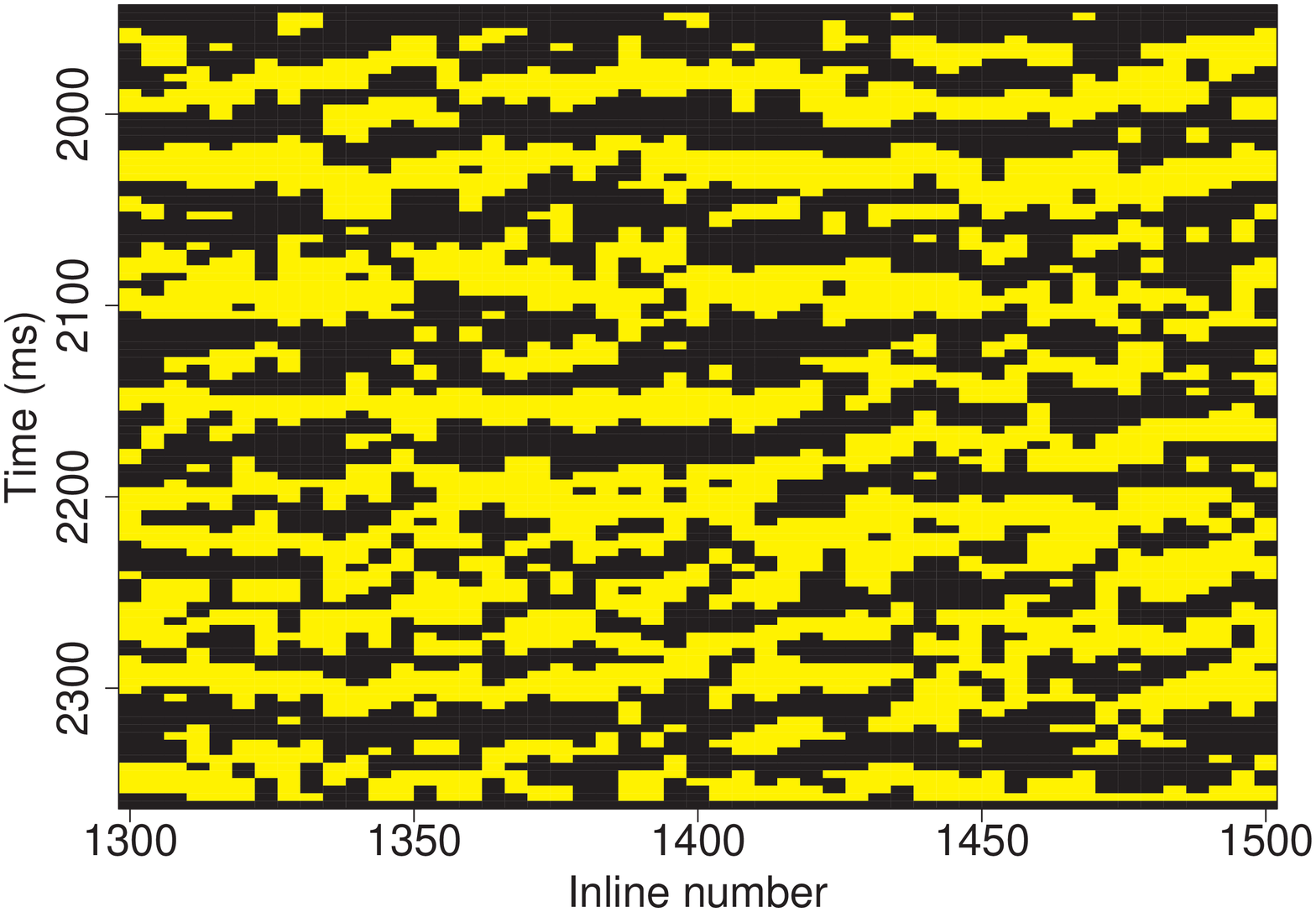}} \\[-0.3cm]
\makebox[6.5cm]{\includegraphics[width=5cm,angle=-90]{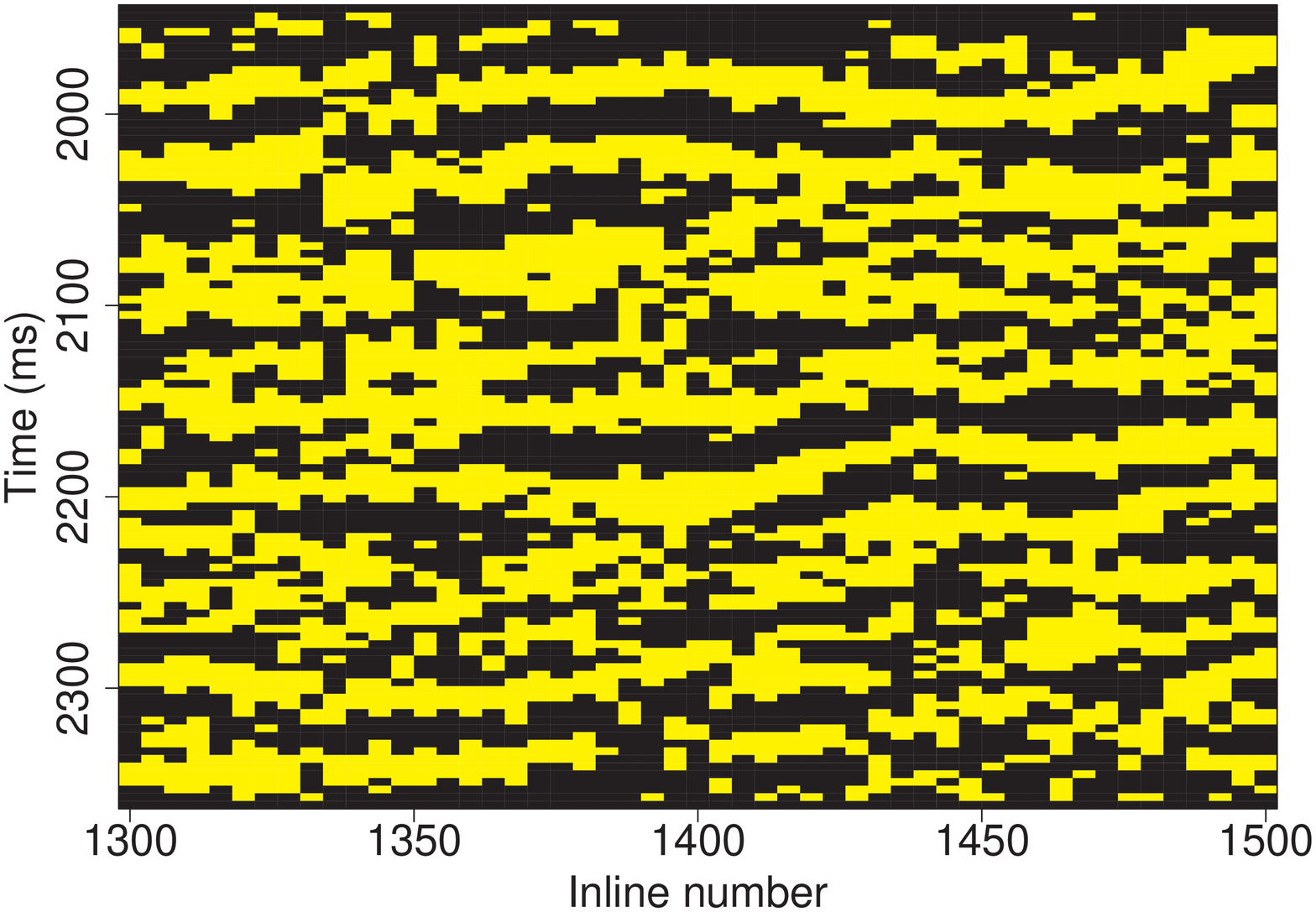}} &
\makebox[6.5cm]{\includegraphics[width=5cm,angle=-90]{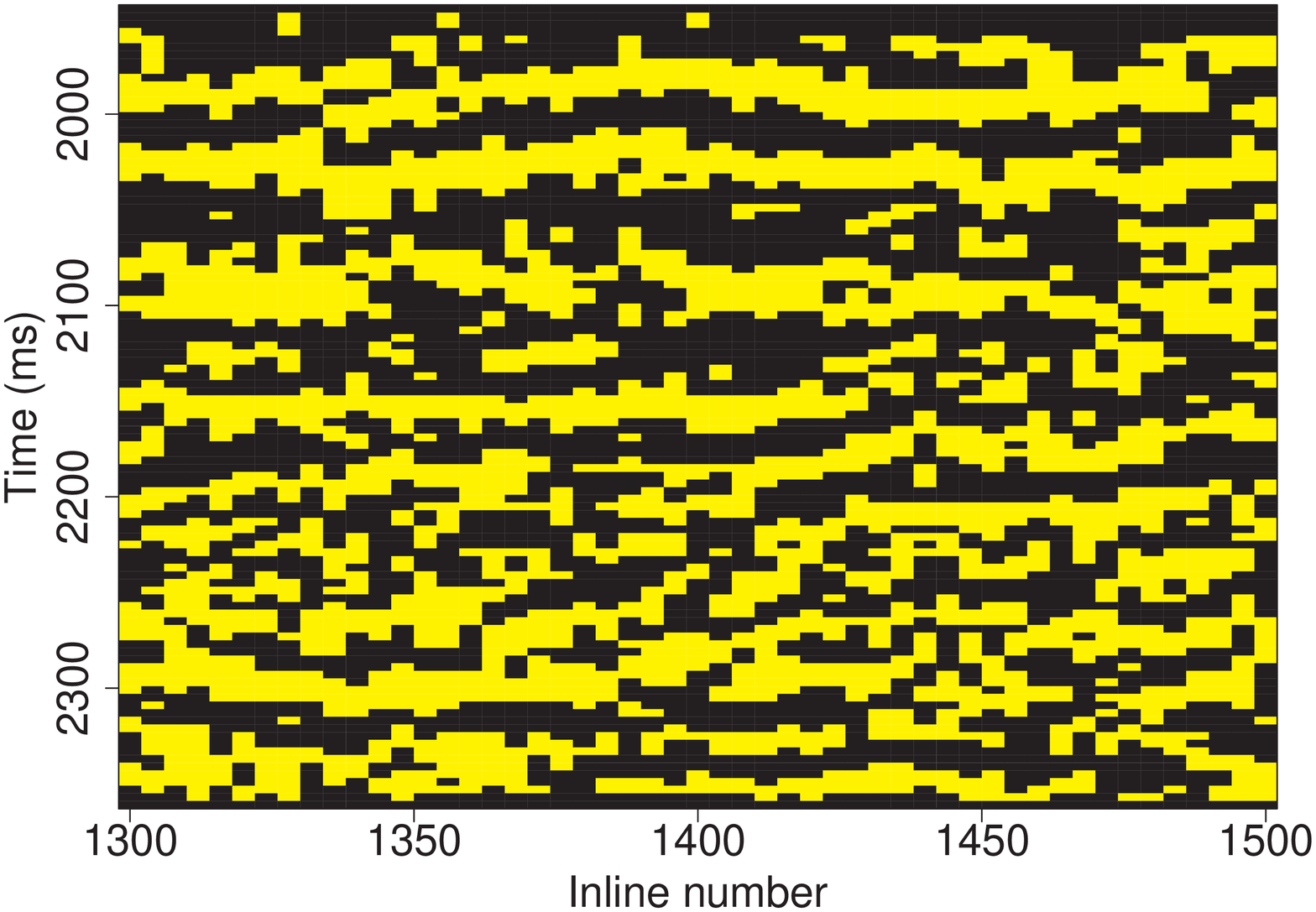}} \\[-0.3cm]
\makebox[6.5cm]{\includegraphics[width=5cm,angle=-90]{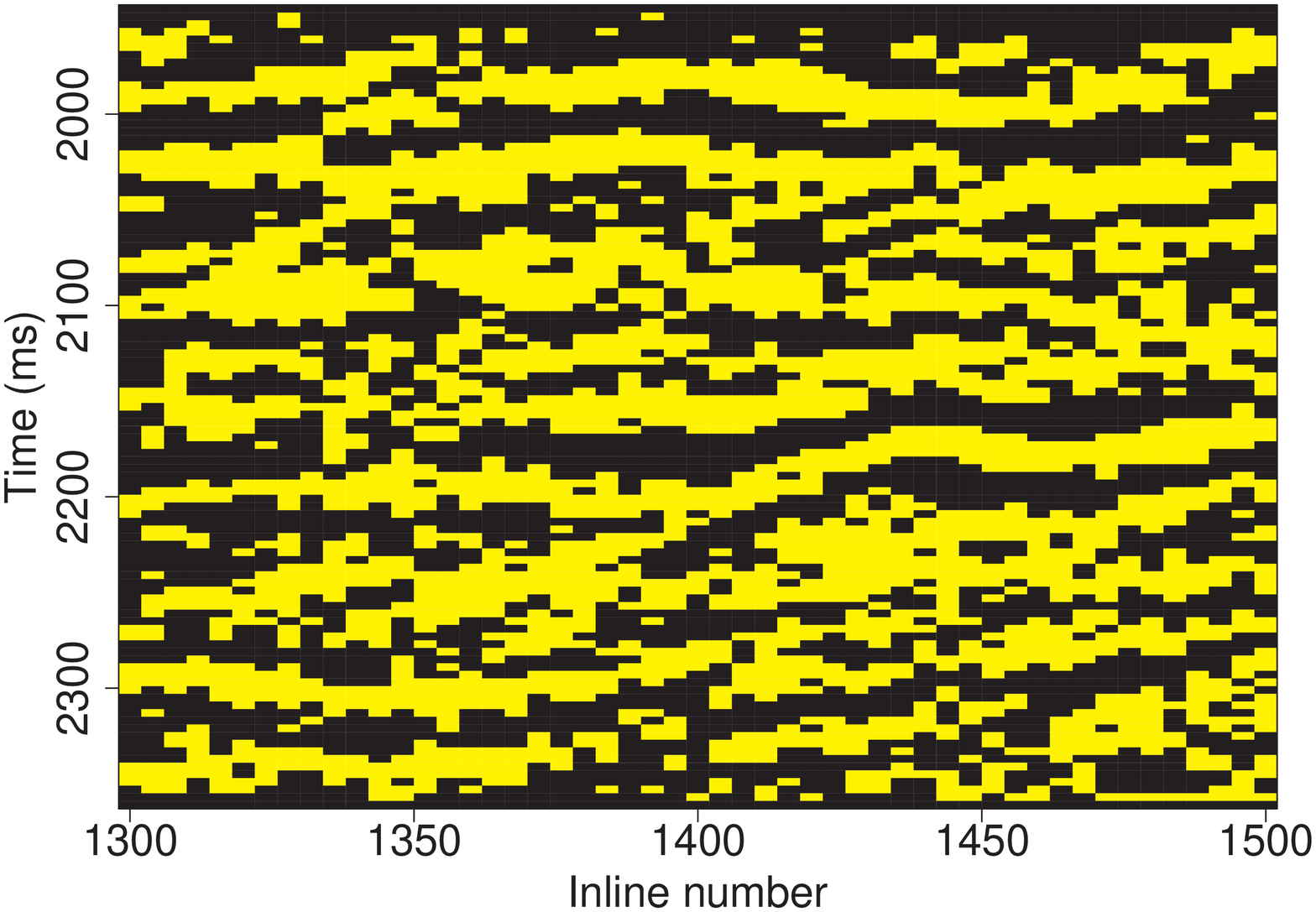}} &
\makebox[6.5cm]{\includegraphics[width=5cm,angle=-90]{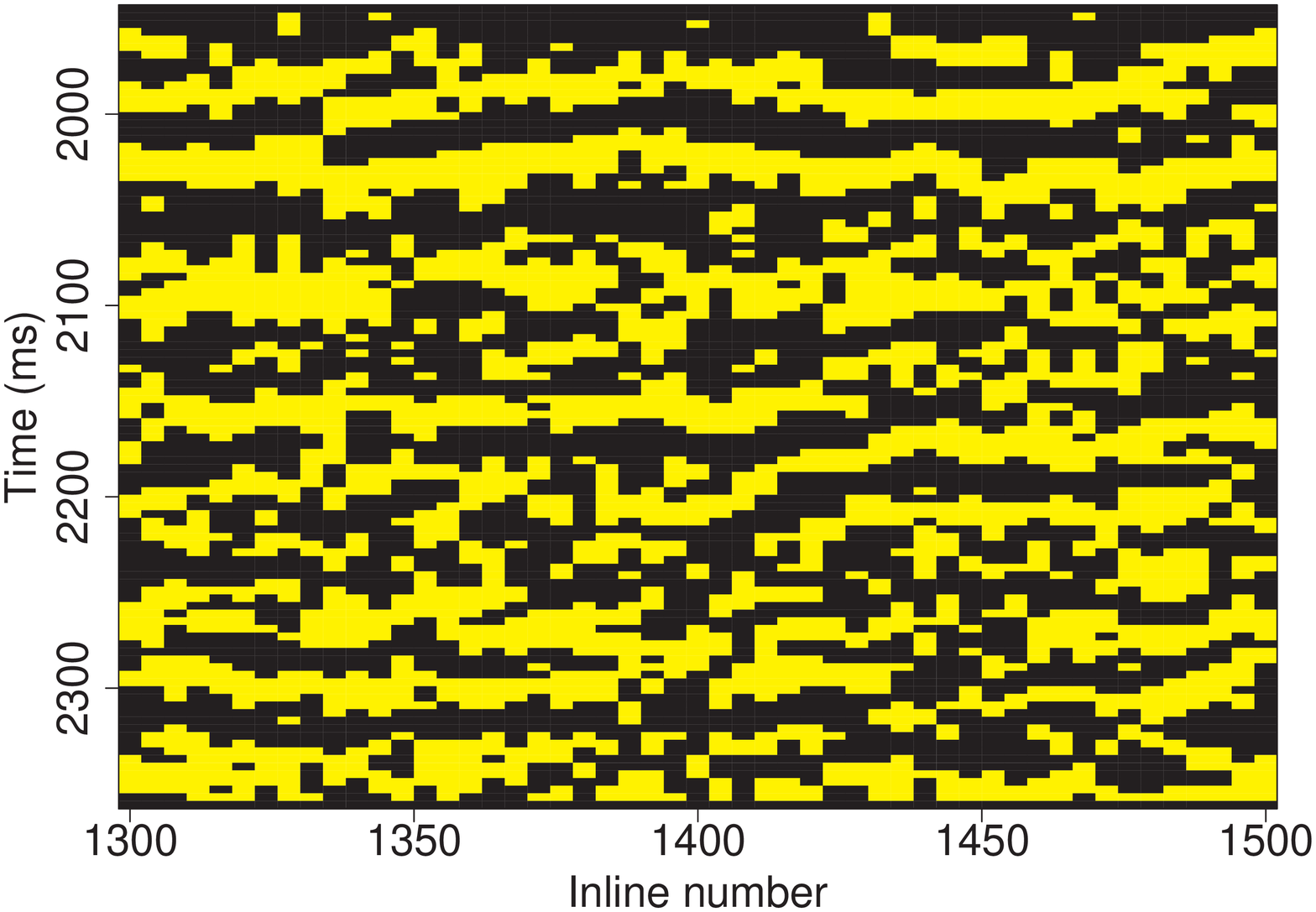}}
\end{tabular}
\end{center}
\caption{\label{fig:realPosterior}The left column shows four independent realisations from the posterior distribution when using the 
Markov mesh prior. The right column shows correspondingly four independent realisations from the posterior distribution when 
using the profile Markov random field prior. Black and yellow represent shale and oil sand, respectively.}
\end{figure}
The left and right columns show four realisations from each of the two posteriors. Realisations from the posterior when using 
the Markov mesh prior are shown in the left column, whereas the realisations in the right column is based on a model with 
the profile Markov random field prior. The eight realisations are quite similar, but when studying them in more detail one 
can observe that with the Markov mesh prior there seems to be more skewed and curved structures than when using 
the profile Markov random field prior. Since the Markov mesh prior has much 
larger neighborhoods than the profile Markov random field prior, this is 
not really surprising. With larger neighborhoods, and corresponding larger cliques, it becomes possible for the model to 
identify skewed and curved structures.

The upper row in Figure \ref{fig:marginalPosterior} 
\begin{figure}
\begin{center}
\begin{tabular}{@{}cc@{}c@{}}
\makebox[6.5cm]{\includegraphics[width=5cm,angle=-90]{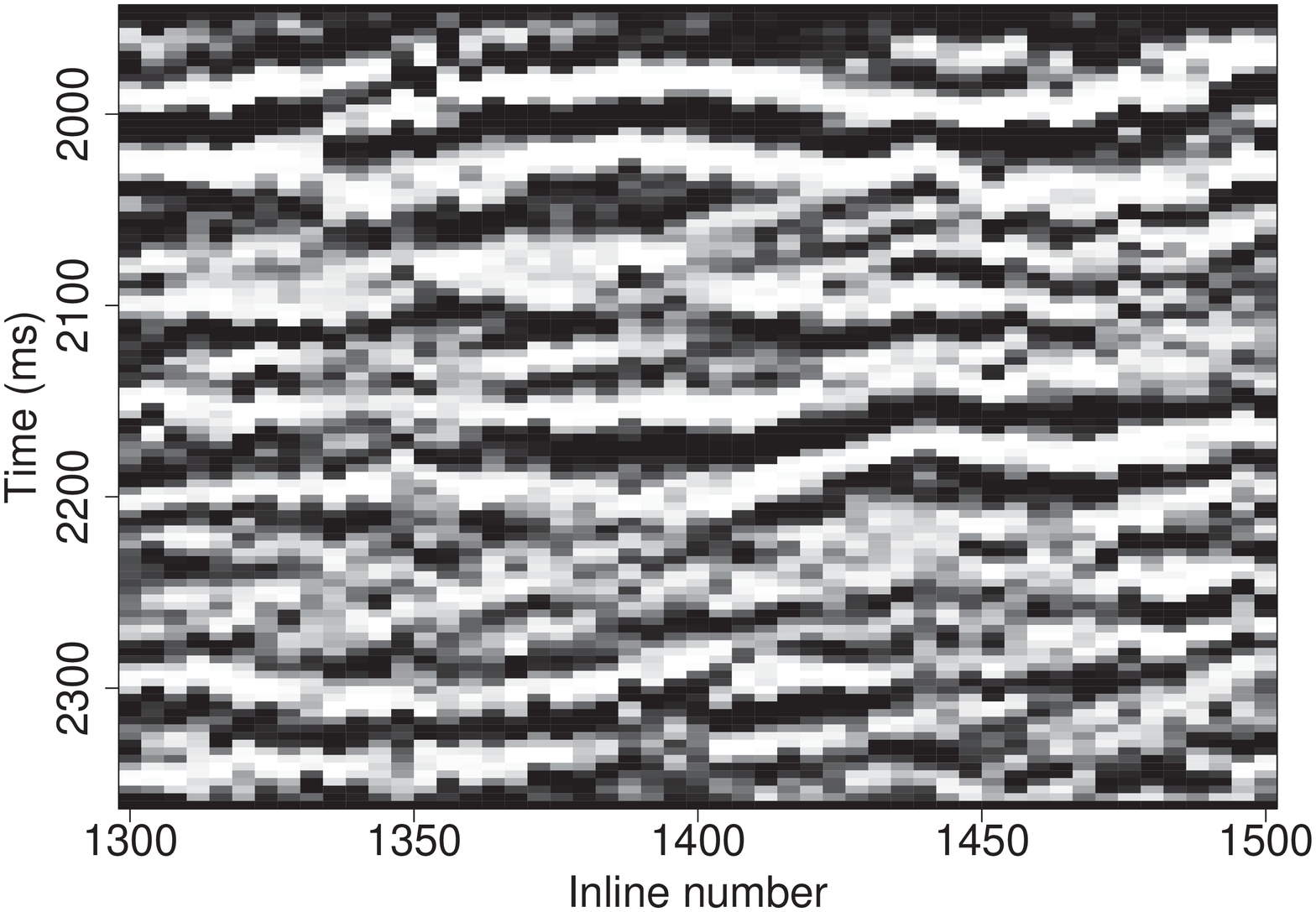}} &
\makebox[6.5cm]{\includegraphics[width=5cm,angle=-90]{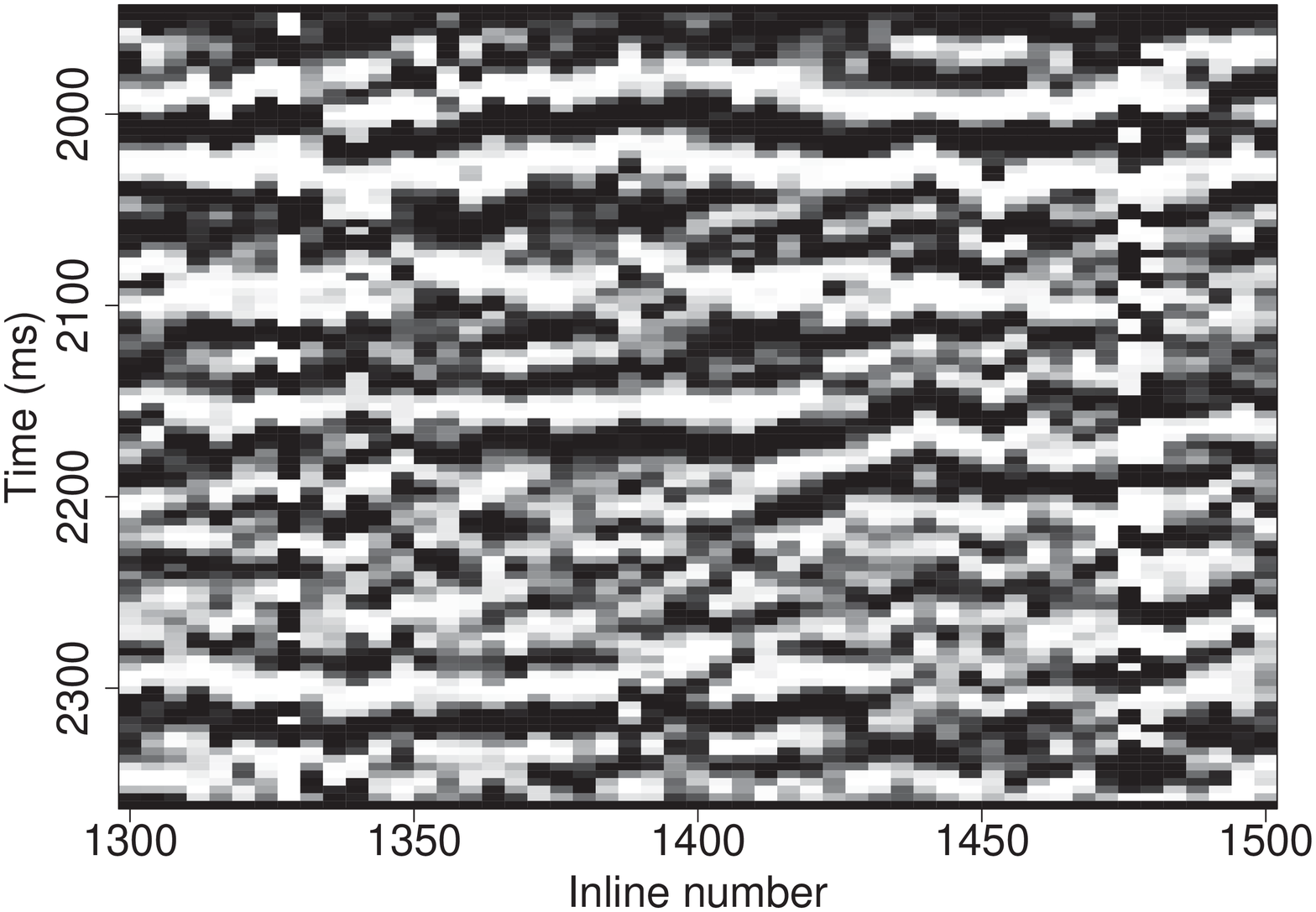}} &
\makebox[6.0cm]{\includegraphics[width=5cm,angle=-90]{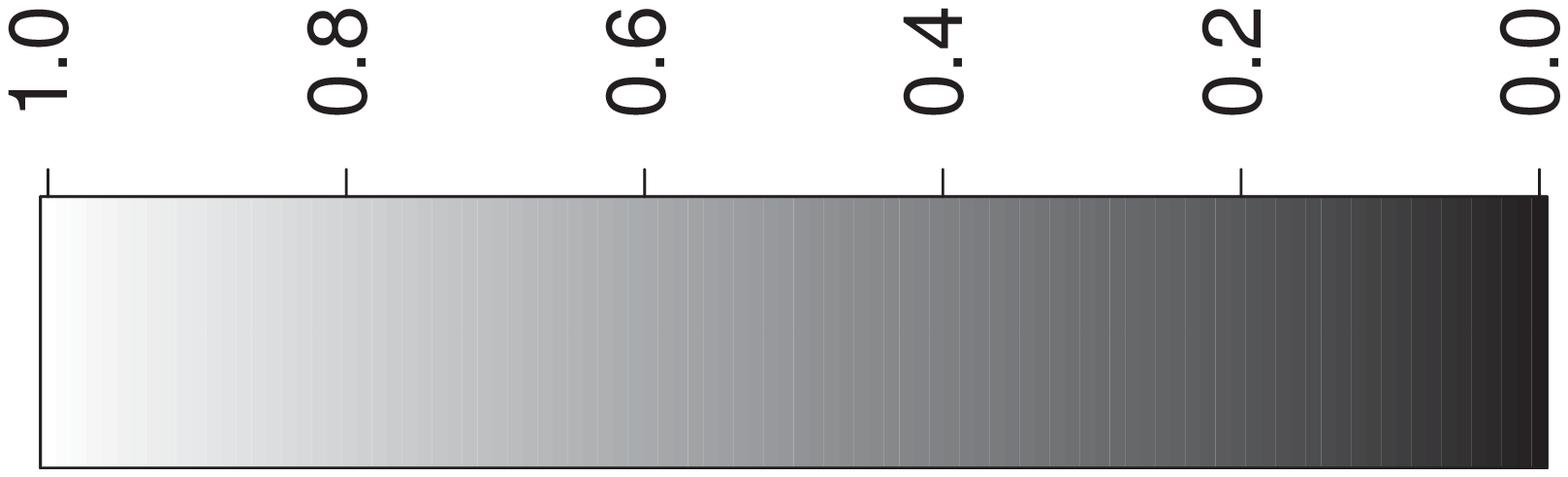}}\\[-0.4cm]
\makebox[6.5cm]{\includegraphics[width=5cm,angle=-90]{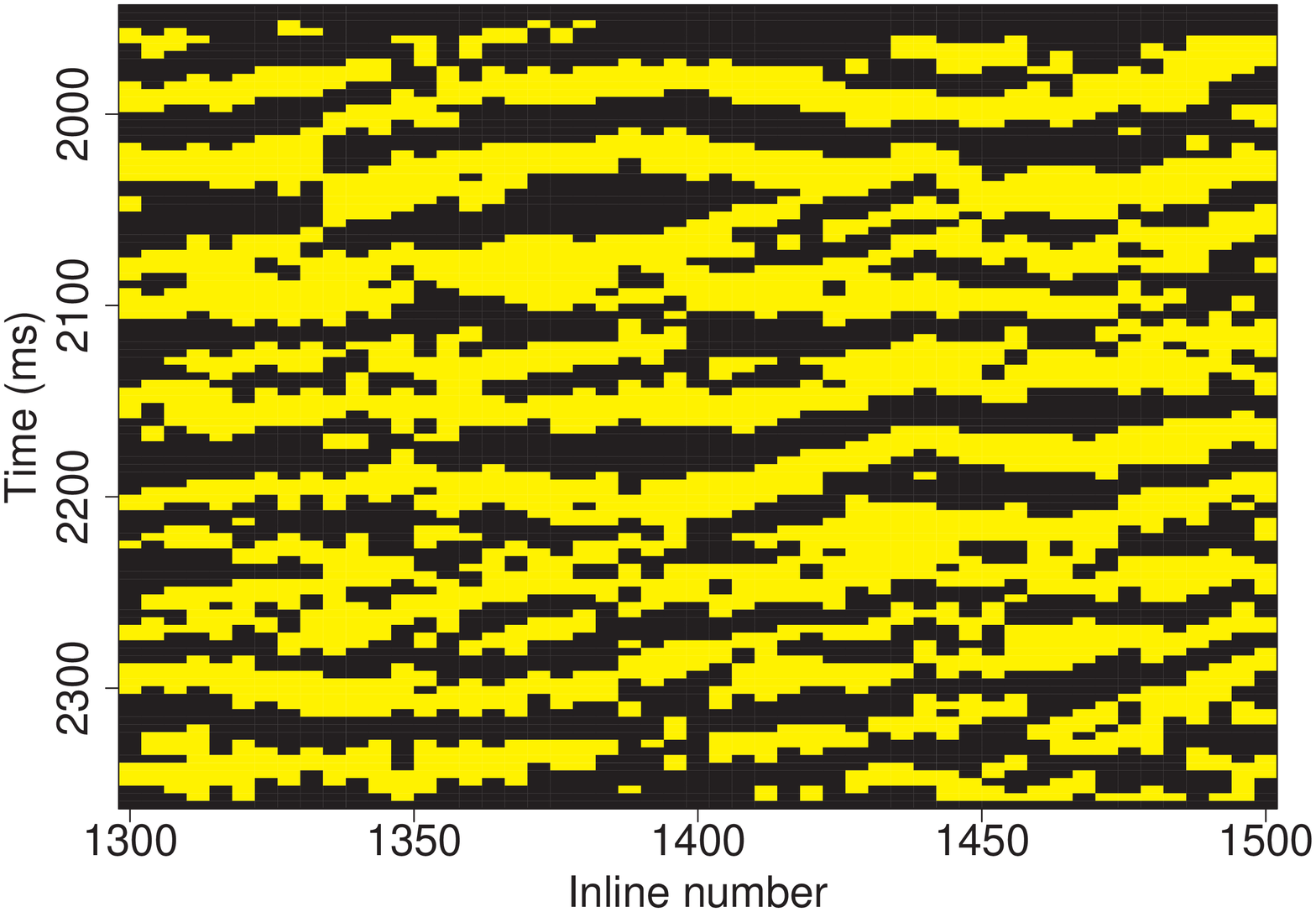}} &
\makebox[6.5cm]{\includegraphics[width=5cm,angle=-90]{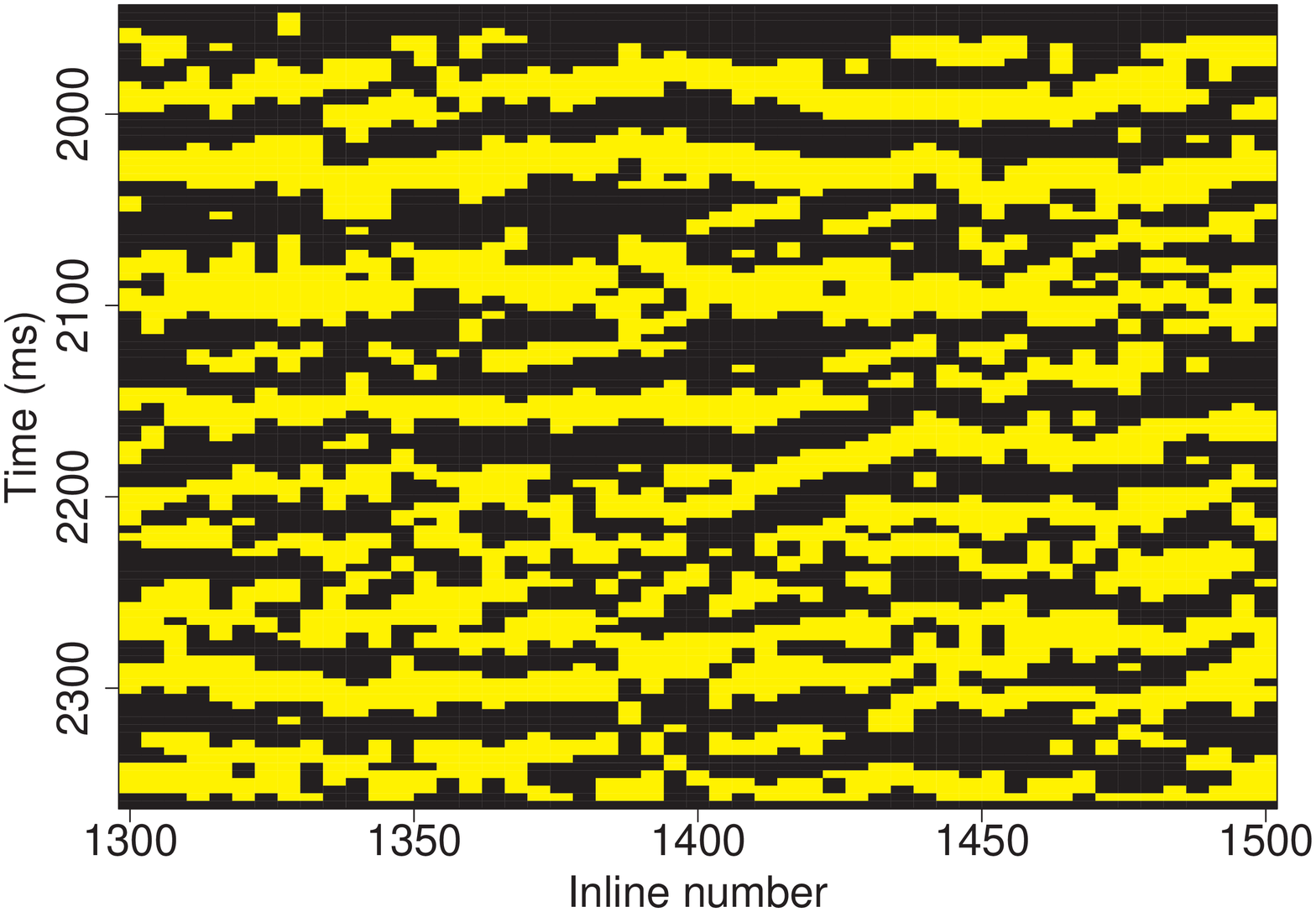}} &
\\[-0.4cm]
\hspace*{0.2cm}\makebox[6.5cm]{\includegraphics[width=5cm,angle=-90]{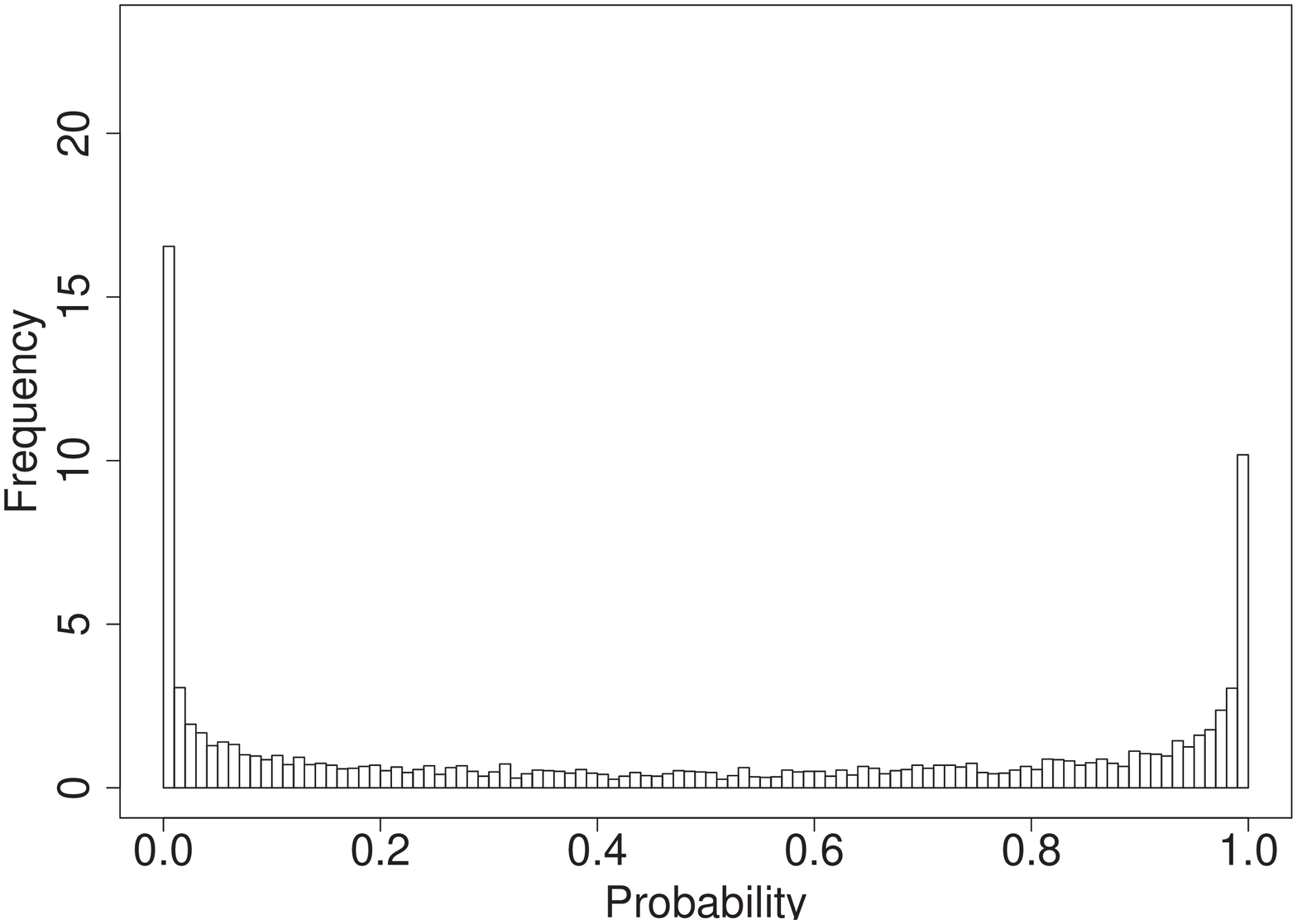}} \hspace*{0.06cm}&
\hspace*{0.2cm}\makebox[6.5cm]{\includegraphics[width=5cm,angle=-90]{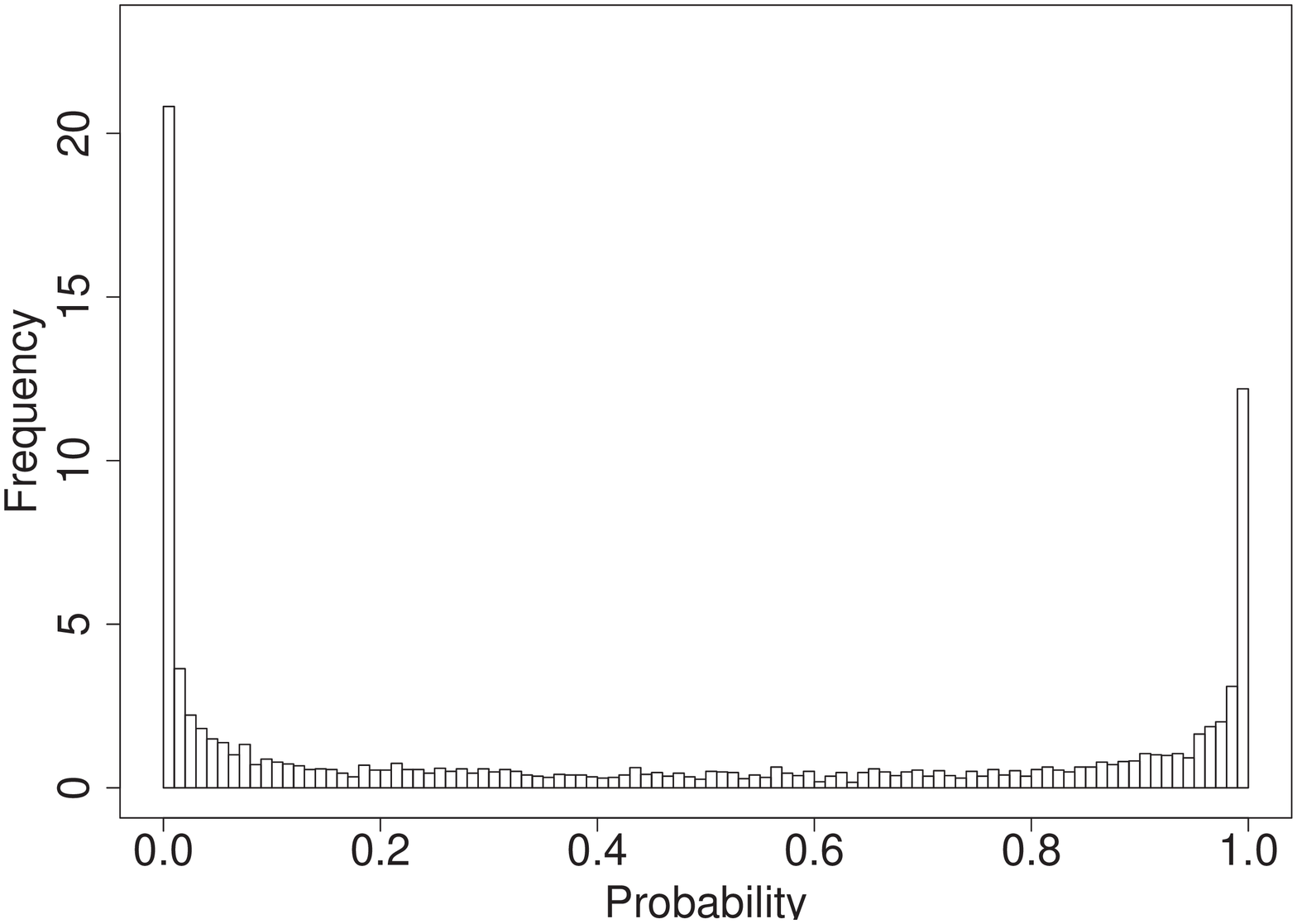}} \hspace*{0.06cm}&
\end{tabular}
\end{center}
\caption{\label{fig:marginalPosterior}Upper row: Estimated posterior marginal probabilities for oil sand when using the 
Markov mesh (left) and the profile Markov random field (middle) priors. The colour scale is shown in the rightmost plot.
Middle row: Estimated marginal posterior mode for each node. Black and yellow represent shale and oil sand, 
respectively. Lower row: Probability histograms of estimated posterior marginal probabilities when using the 
Markov mesh (left) and the profile Markov random field (right) priors.}
\end{figure}
shows the result of estimating in each node the marginal posterior 
probability of oil sand. In each node the probability is estimated as the fraction of the realisations where the node
has oil sand. Again the left and right images are results when using the Markov mesh and profile Markov random field 
priors, respectively. The two probability maps are similar, but somewhat more continuity of skewed and curved 
high probability areas can be observed when using the Markov mesh prior. In the middle row of Figure \ref{fig:marginalPosterior}
the probabilities in the upper row is rounded to the nearest integer to get an estimate of the most probable
lithology/fluid class in each node. Again we can observe somewhat more continuity of skewed and curved oil sand areas when 
using the Markov mesh prior. The histograms in the lower row of Figure \ref{fig:marginalPosterior} are
simply probability histograms of the 
estimated marginal posterior probabilities shown in the upper row of the same figure. We can observe that when using 
the profile Markov random field prior, somewhat more marginal posterior probabilities are close to zero and one than when using 
the Markov mesh prior. 

To study the marginal probabilities a little more we have chosen three traces, or columns, $j=15$, $30$ and $45$, and 
in Figure \ref{fig:traces} plotted the marginal probabilities.
\begin{figure}
\begin{center}
\begin{tabular}{cccccc}
\makebox[4.9cm]{~} &
\makebox[3.5cm]{\includegraphics[width=10cm,angle=-90]{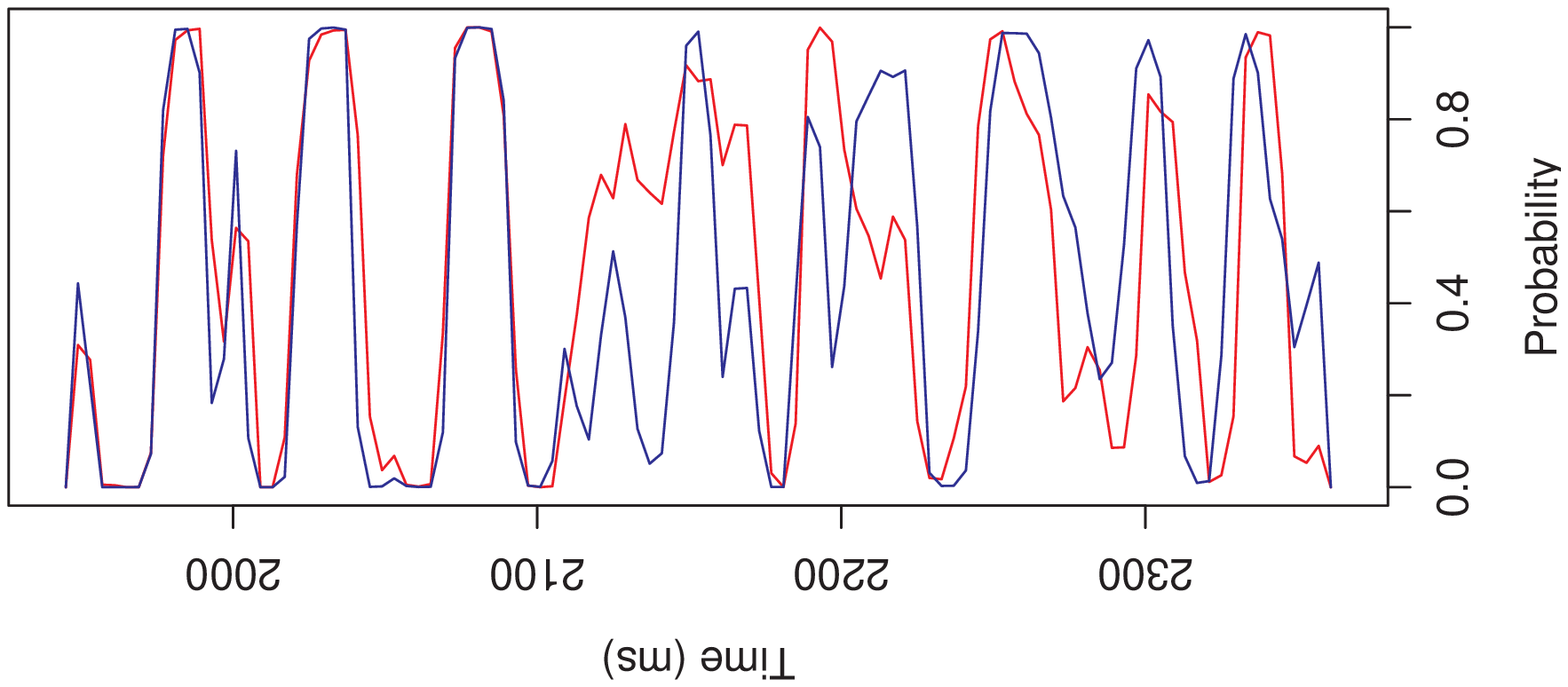}} &
~~~~&
\makebox[3.5cm]{\includegraphics[width=10cm,angle=-90]{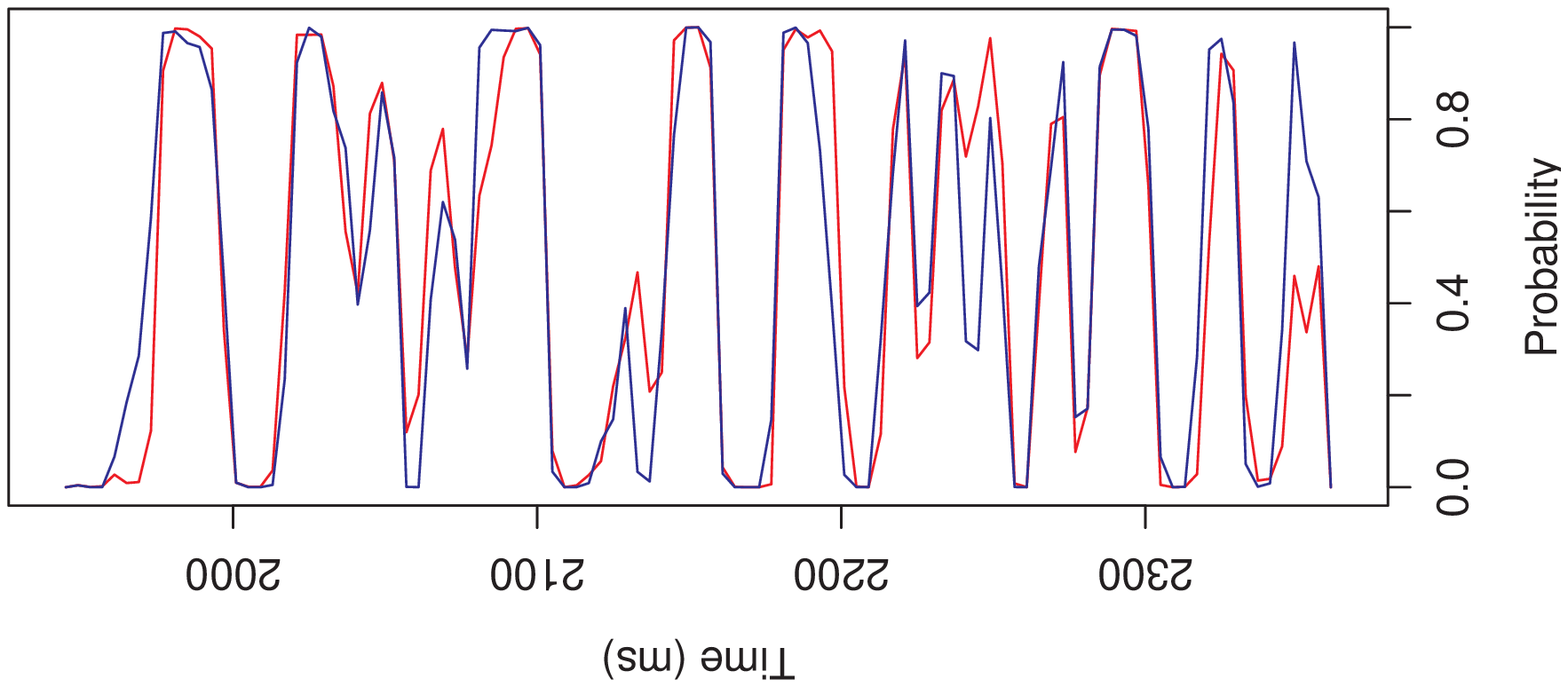}} &
~~~~&
\makebox[3.5cm]{\includegraphics[width=10cm,angle=-90]{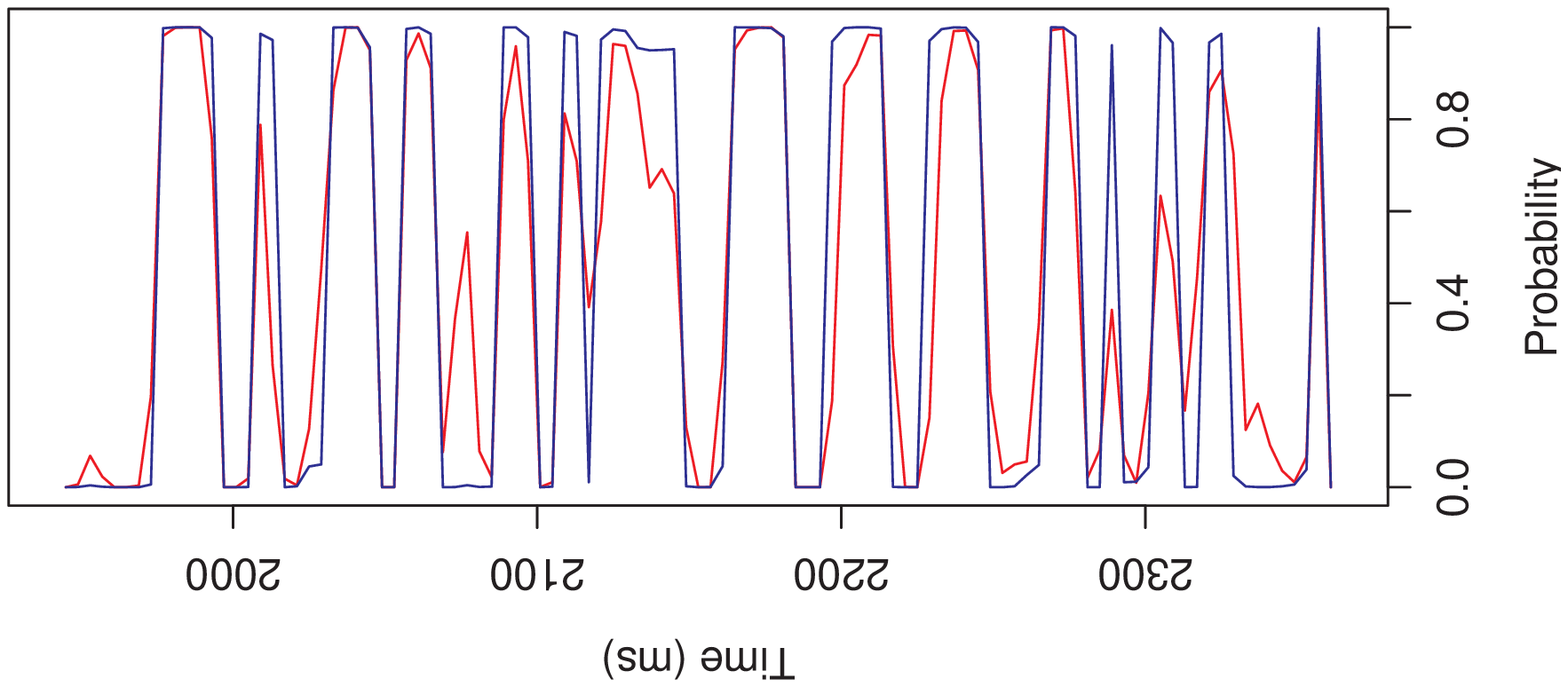}}
\end{tabular}
\end{center}
\caption{\label{fig:traces}Estimated marginal probabilities for oil sand in three traces. Results when using 
the Markov mesh and profile Markov random field priors are shown by red and blue lines, respectively.
The left, middle and right plots show the marginal probabilities
in traces $j=15$, $j=30$ and $j=45$, respectively.}
\end{figure}
The estimated posterior marginal probabilities when using the Markov mesh and profile Markov random field priors
are plotted in red and blue, respectively. More than in Figure \ref{fig:marginalPosterior} we can here see 
how close the two posterior probabilities are for most of the nodes. In a few of the nodes, however, the difference
is quite clear. 

The continuity of oil sand is very important for fluid flow in a petroleum reservoir. We can get some 
understanding of how the prior influences this continuity by studying Figure \ref{fig:marginalPosterior}, but 
to study the continuity in more detail we need to summarise how this continuity is in each posterior realisation. 
To do this, we have manually picked four nodes with very high posterior probability for oil sand both 
when using the Markov mesh and 
the profile Markov random field priors. These four nodes are marked with a red bullet in Figure \ref{fig:contact}, one 
row for each of the four chosen nodes.
\begin{figure}
\begin{center}
\begin{tabular}{@{}cc@{}}\\[-2.0cm]
\makebox[6.5cm]{\includegraphics[width=5cm,angle=-90]{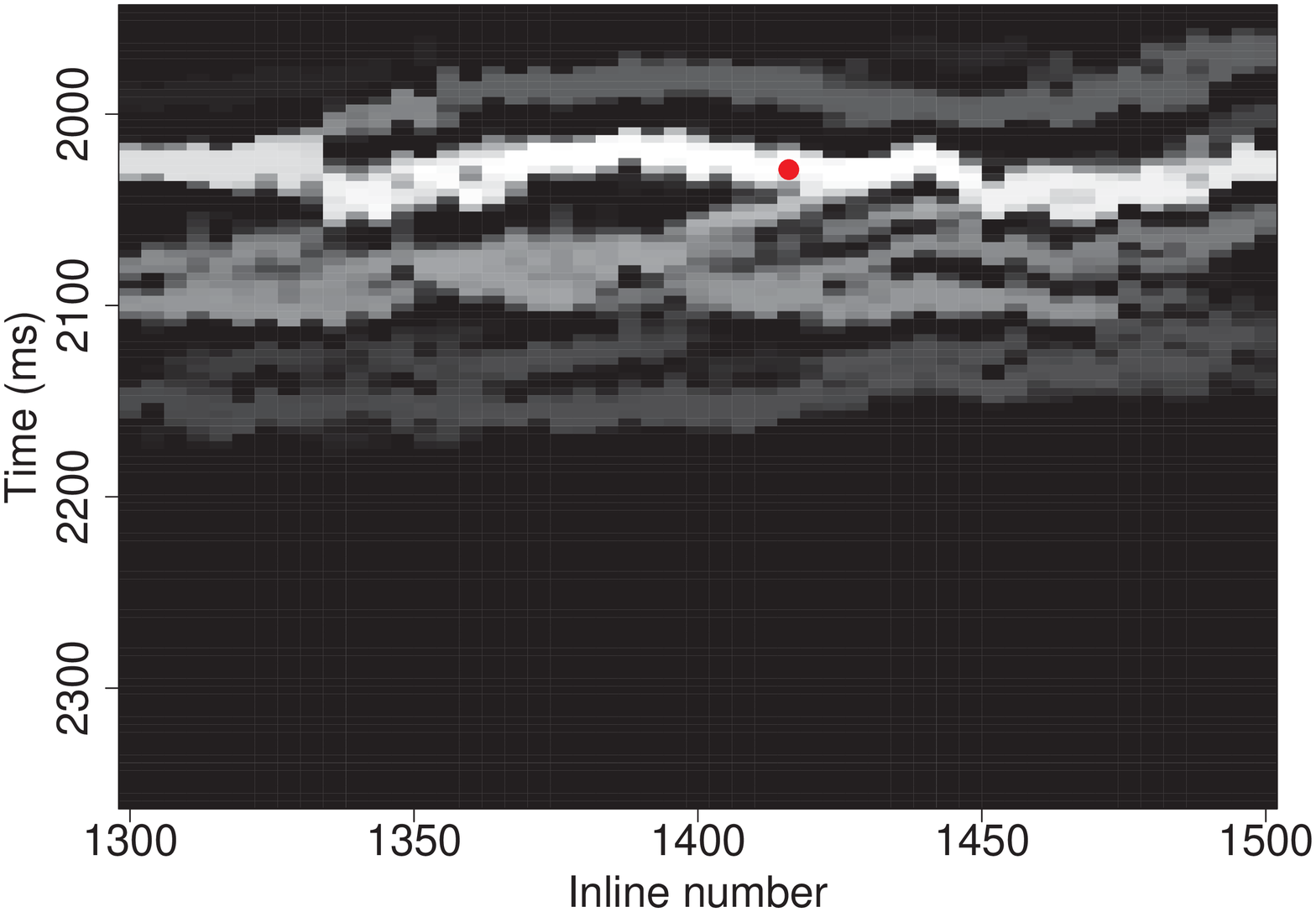}} &
\makebox[6.5cm]{\includegraphics[width=5cm,angle=-90]{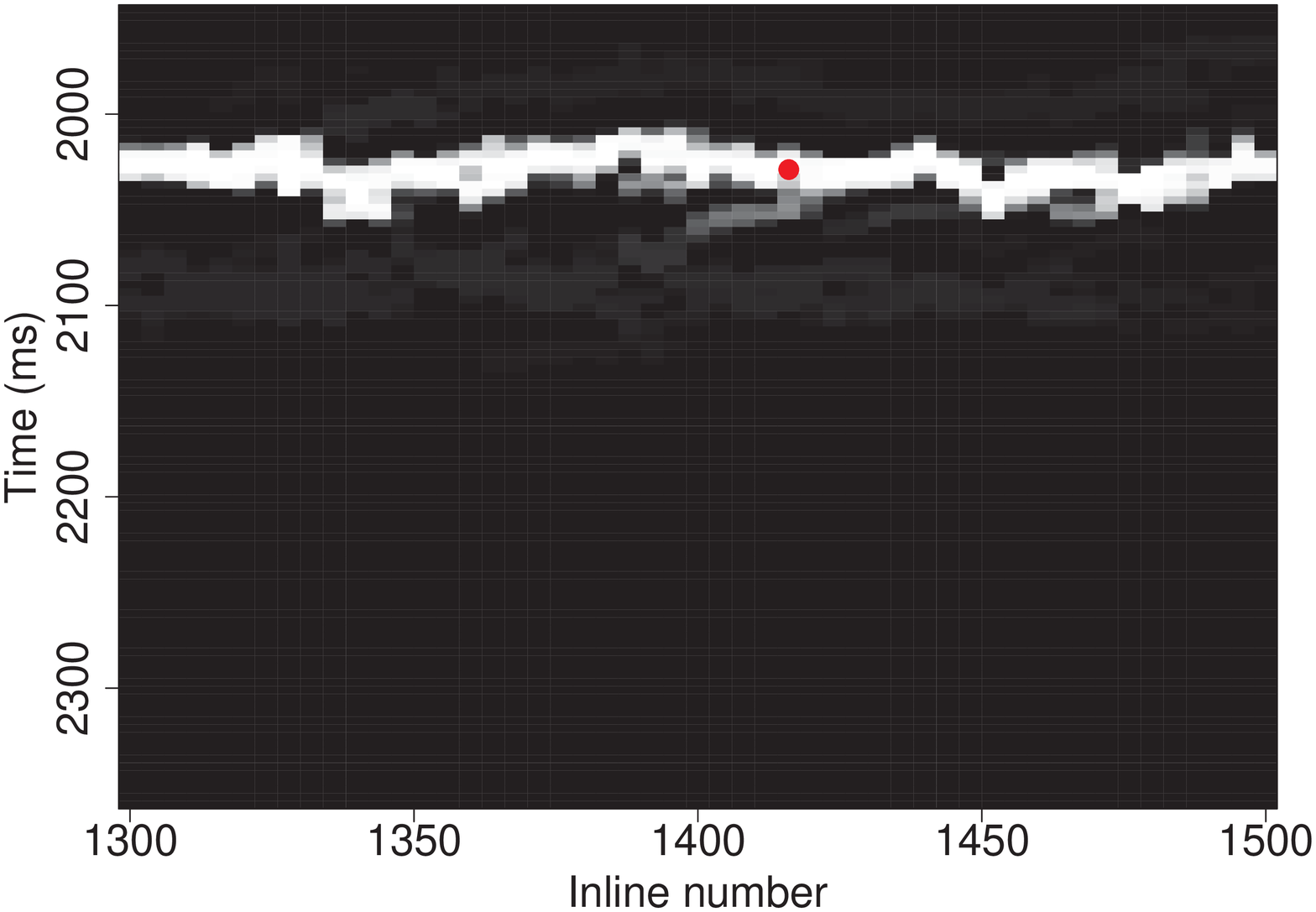}} \\[-0.3cm]
\makebox[6.5cm]{\includegraphics[width=5cm,angle=-90]{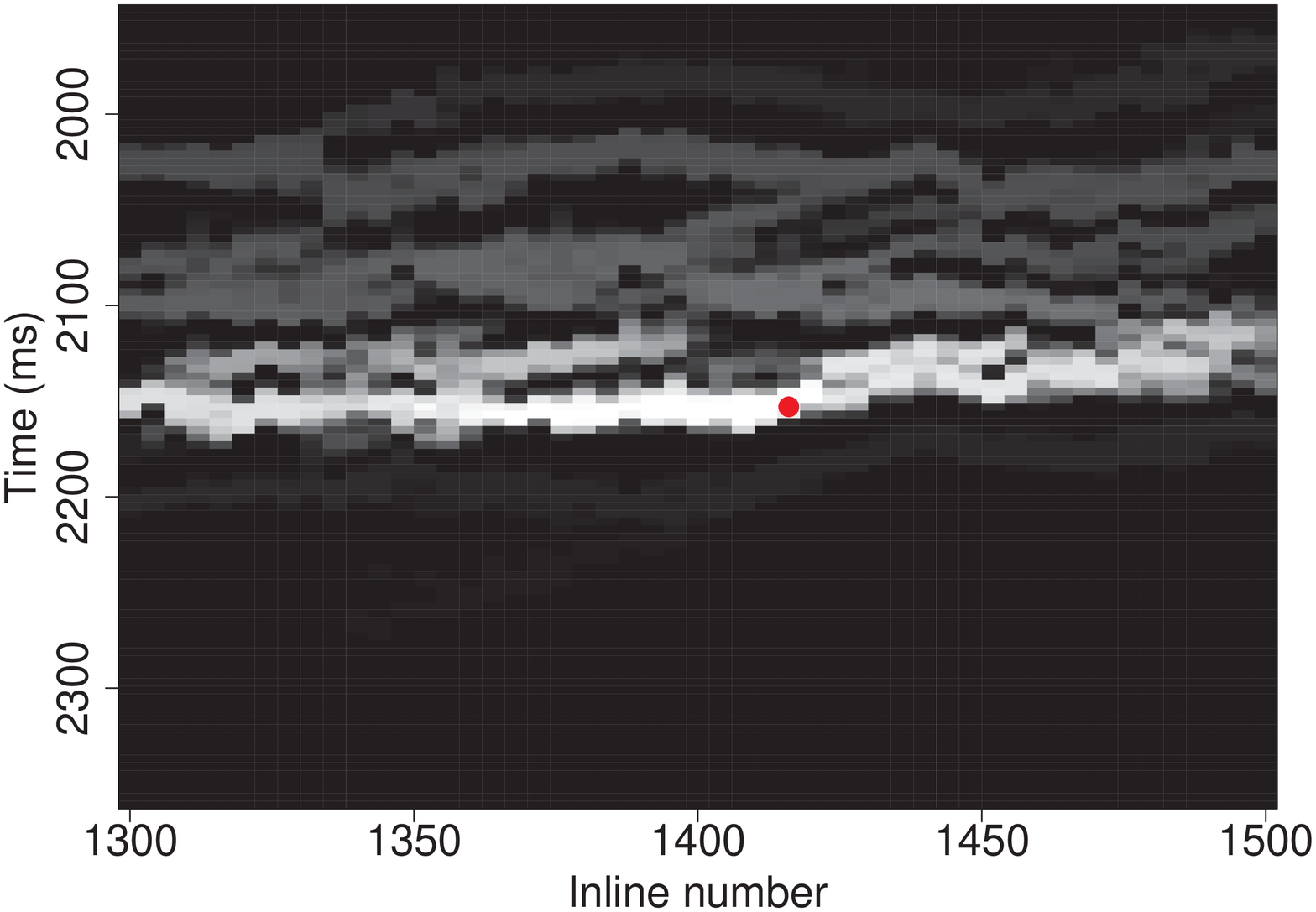}} &
\makebox[6.5cm]{\includegraphics[width=5cm,angle=-90]{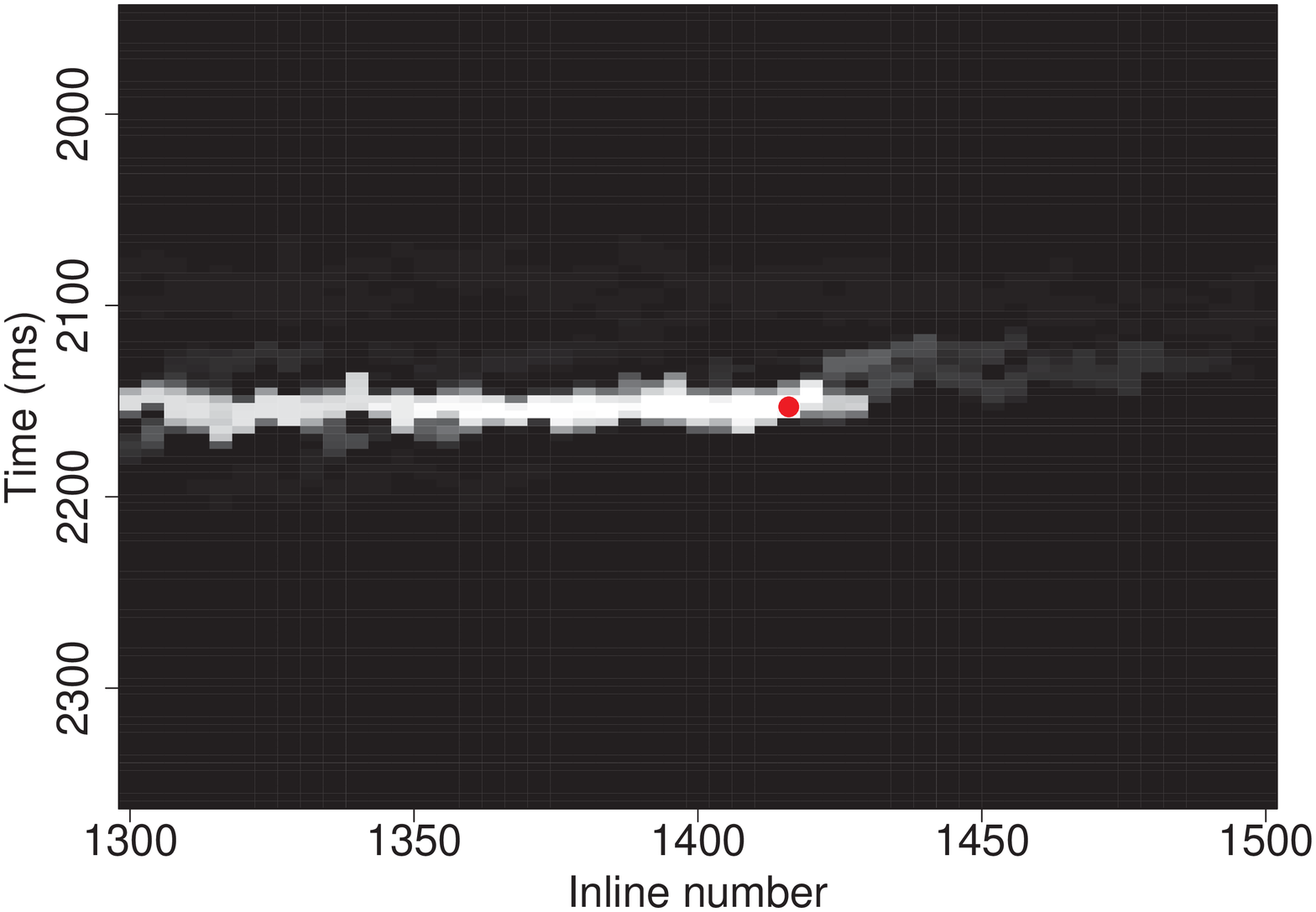}}\\[-0.3cm]
\makebox[6.5cm]{\includegraphics[width=5cm,angle=-90]{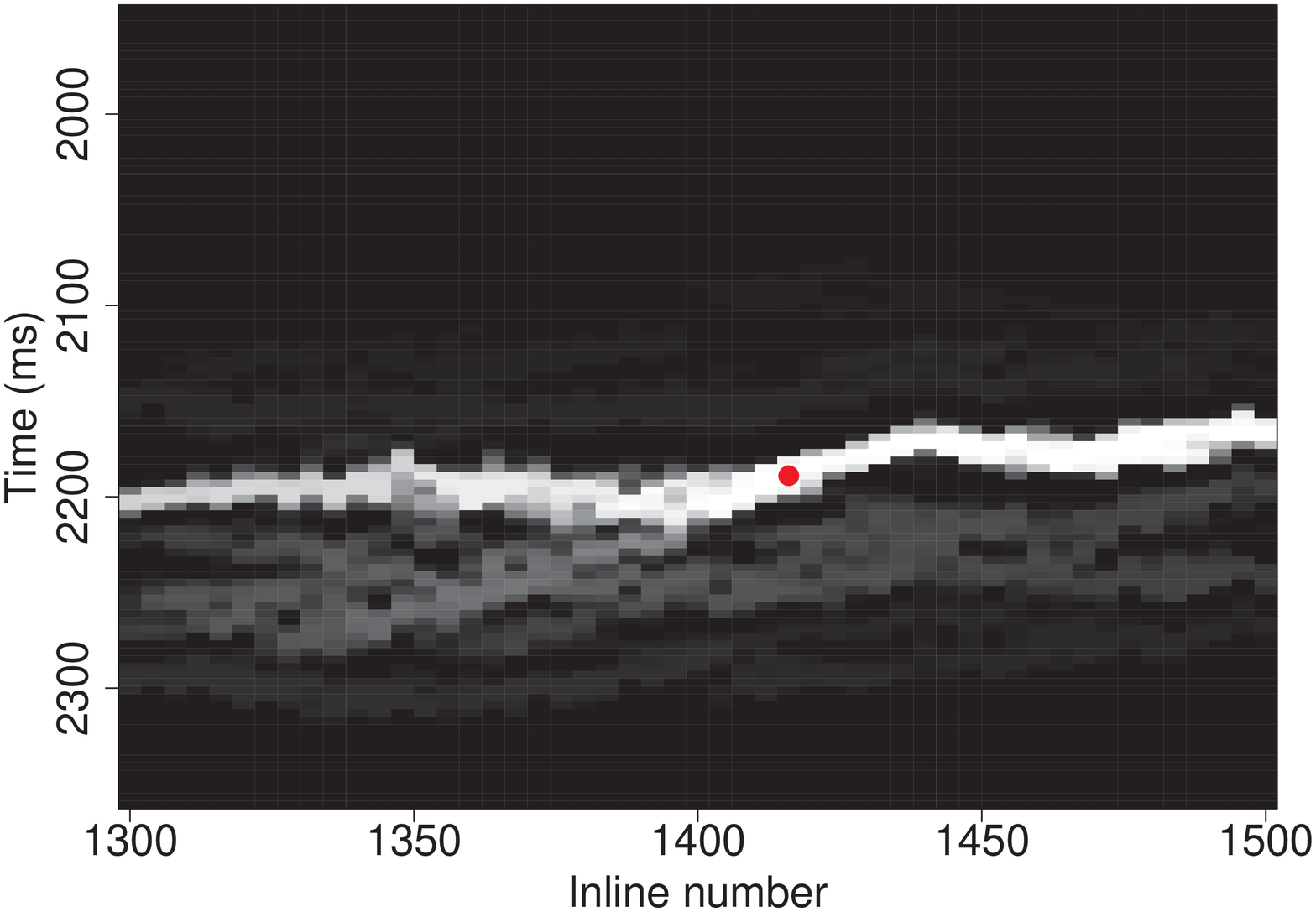}} &
\makebox[6.5cm]{\includegraphics[width=5cm,angle=-90]{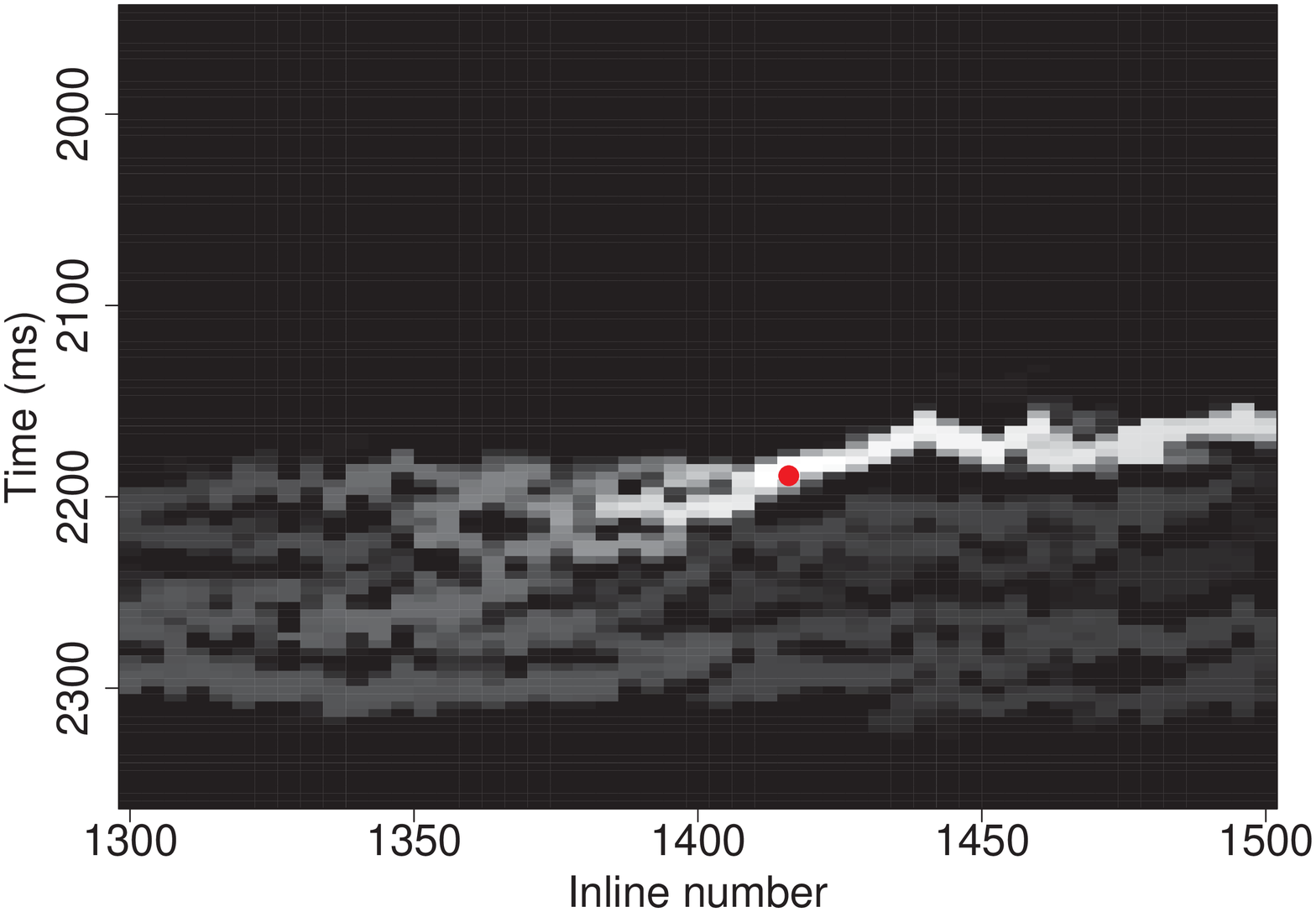}}\\[-0.3cm]
\makebox[6.5cm]{\includegraphics[width=5cm,angle=-90]{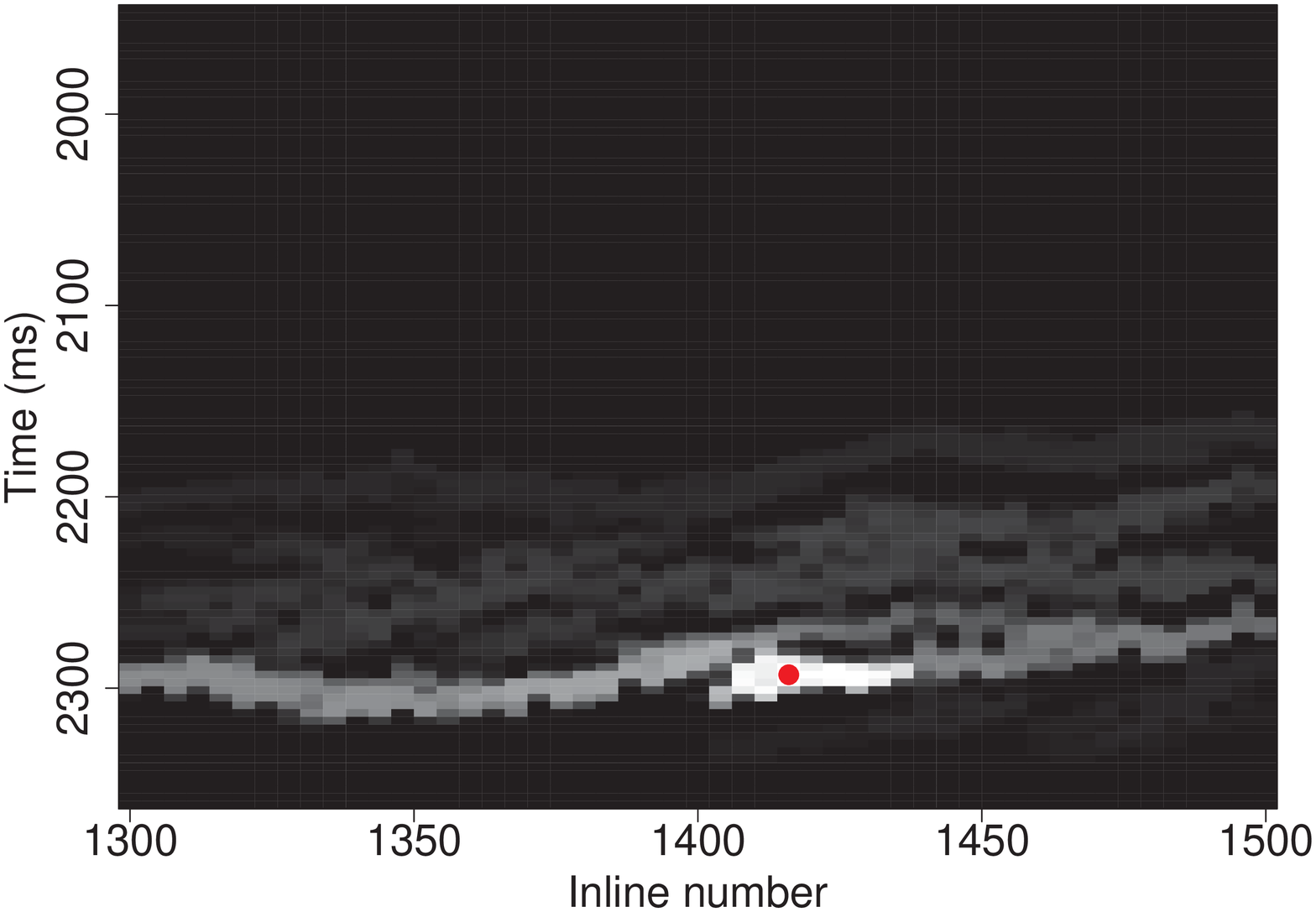}} &
\makebox[6.5cm]{\includegraphics[width=5cm,angle=-90]{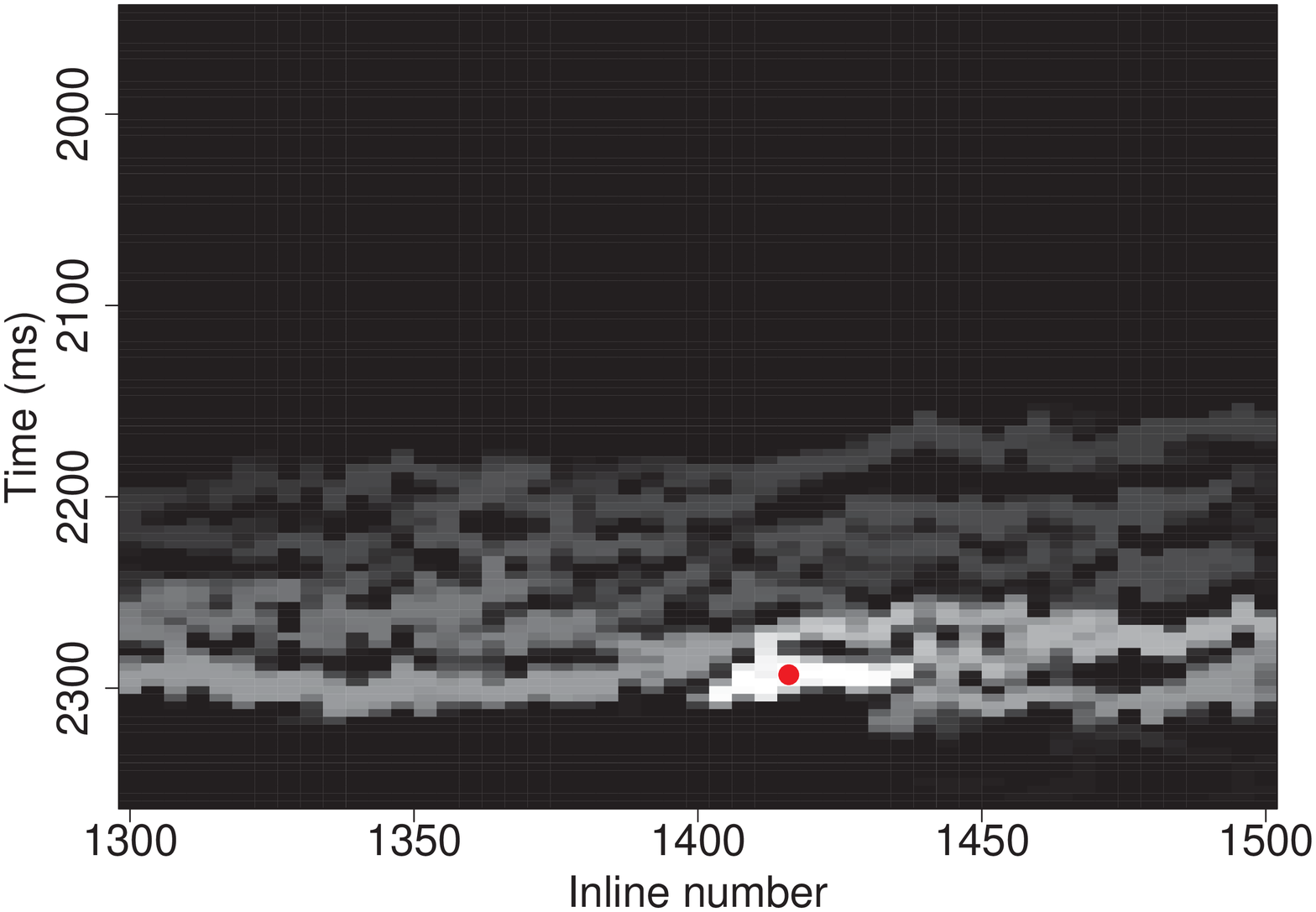}}
\end{tabular}
\end{center}
\caption{\label{fig:contact}For each node, estimated posterior probabilities that there is contact (via other
oil sand nodes) between this node and the node marked with a red filled circle. The plots in the left and right columns
are when using the Markov mesh and the profile Markov random field priors, respectively. The colour scale is as defined
by the legend in Figure \ref{fig:marginalPosterior}.}
\end{figure}
For each of these four nodes and for each posterior realisation we identified all other nodes with oil sand which through other
oil sand nodes had contact with the chosen node. Thereby we could estimate the posterior probability 
that any node was in contact with the chosen node as the fraction of the realisations where this occurred. The resulting
probability plots are shown in Figure \ref{fig:contact}. The left and right columns are again the results when using 
the Markov mesh and the profile Markov random field priors, respectively. In the three upper rows we can see a lot more
continuity in the posterior realisations when using the Markov mesh prior than when using the profile Markov random field 
prior. In the lower row the situation is for some reason reversed. To study this type of continuity more generally, not only 
for the four hand picked nodes used in Figure \ref{fig:contact}, we finally repeat the exercise of finding all nodes 
in a realisation with oil sand connected to a particular node, but now the particular node is sampled at random among all 
nodes with oil sand. For each realisation and each particular node we find the number of oil sand nodes connected to 
the particular node. In Figure \ref{fig:connectivityArea} we show the 
\begin{figure}
\begin{center}
\begin{tabular}{c}
\includegraphics[width=6cm,angle=-90]{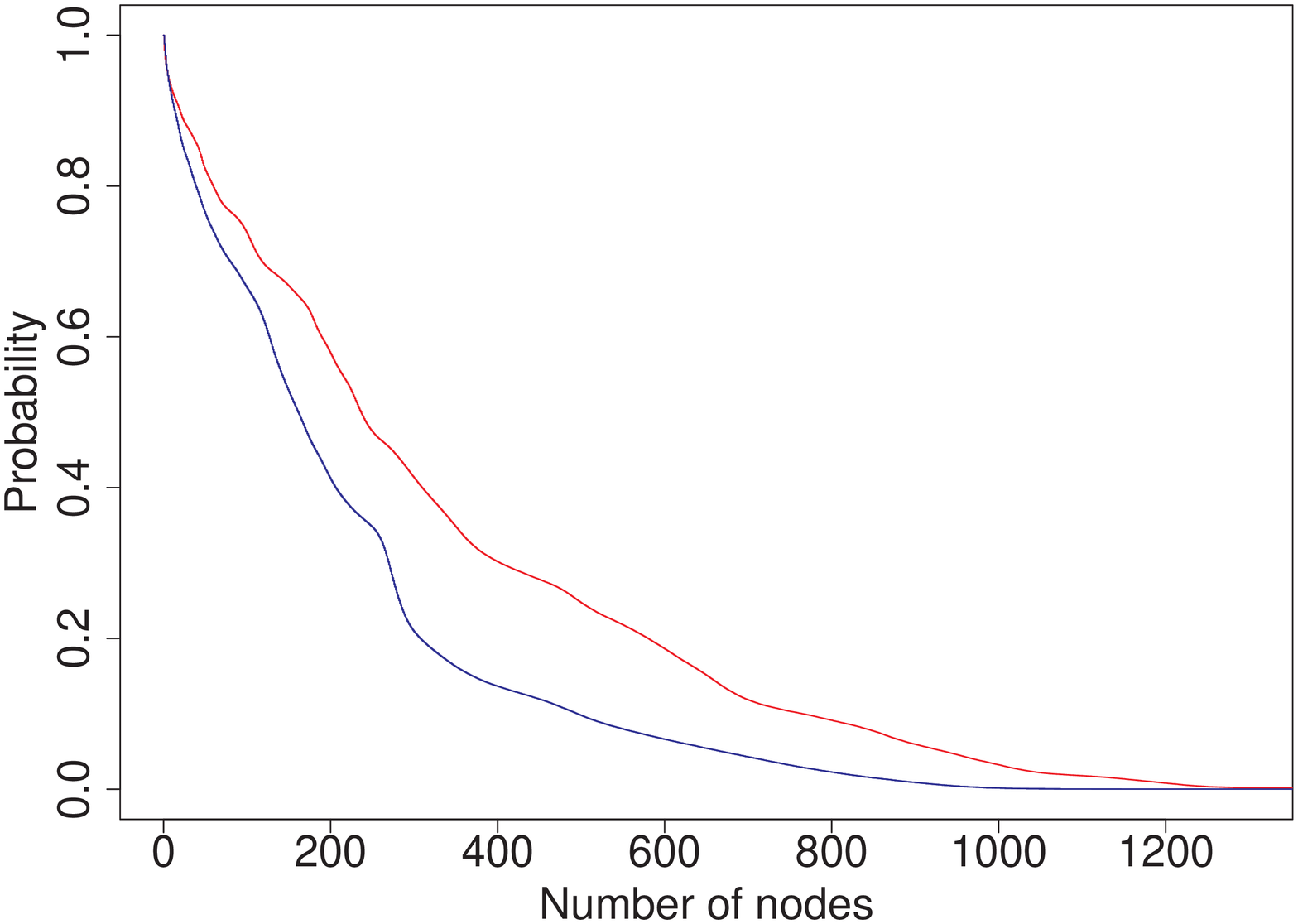}
\end{tabular}
\end{center}
\caption{\label{fig:connectivityArea}As a function of a number of nodes $\eta$, estimated posterior probability 
  for a random oil sand node to be connected (via other oil sand nodes) to at least $\eta$ other oil sand nodes. The $\eta$ is along
the $x$-axis and the estimated probability is along the $y$-axis. Results when using the Markov mesh and profile
Markov random field priors are shown in red and blue, respectively.}
\end{figure}
resulting estimated posterior probabilities for the randomly chosen particular node to be connected to more than $\eta$ 
other oil sand nodes, as a function of $\eta$. The red and blue curves are the results when using the Markov mesh and 
the profile Markov random field priors, respectively. We see that the curve related to the Markov mesh prior lies
consistently clearly above the curve related to the profile Markov random field prior, showing that the Markov mesh prior produces
more posterior continuity of oil sand than the profile Markov random field prior.

 \section{Discussion}
In this article we have, for a particular seismic data set in the North Sea and two particular prior models for the 
lateral connectivity of the lithology/fluid classes, 
studied how the prior influences the posterior properties. When focusing on the posterior marginal 
probabilities we found, for most nodes, that the prior had little influence. When focusing on posterior continuity
of oil sand, however, we found that the prior had a quite strong influence on the results. Not 
surprisingly, the prior with the largest neighborhoods produced the largest posterior continuity. When 
evaluating whether to use a simple prior with a small neighborhood or to use a more complicated prior with a 
larger neighborhood one should therefore first decide what posterior properties that are of interest. If the focus
is only on the posterior marginal probabilities, a simple prior is perhaps sufficient. If the focus is on 
fluid flow, however, spatial continuity is crucial and it may be beneficial with a more complicated prior which is better able to 
capture spatial continuity.

 When deciding what prior to use one should also take into account
the computational resources necessary to simulate from the posterior distribution. A more complicated prior 
typically gives a posterior which requires more computation time to explore. With our implementations, sampling 
from the posterior when using the Markov mesh prior required approximately $20$ times more computation time
compared to when using the profile Markov random field prior. However, our implementation of the sampling when 
using the Markov mesh prior was partly in Matlab and partly in C++ and a lot of the computation time here was 
just overhead in the communication between Matlab and C++. Our implementation of the sampling algorithm when 
using the profile Markov random field prior was entirely in Matlab, so we did not have the same overhead in 
this case. If we had implemented also the sampling algorithm when using the Markov mesh prior entirely in 
Matlab we expect this algorithm would have required a factor between $3$ and $5$ more computation time than that for 
the profile Markov random field prior. That the sampling when using the Markov mesh prior requires more computation
time than when using the profile Markov random field prior should come as no surprise, since the Markov mesh prior 
has a much larger neighborhood than the profile Markov random field prior.

 \subsection{Future research}

The study presented in this article is quite limited. The model includes only two lithology/fluid classes and we have considered
only one seismic section. It is of interest also to study the effect of using a prior with a larger neighborhood
when the model represent more than two lithology/fluid classes. The profile Markov random field prior is already 
defined with more than two lithology/fluid classes and the Markov mesh construction used here 
can easily be extended to such a situation. To condition on a seismic cube is also of interest. Then the 
lithology/fluid classes need to be represented on a three dimensional lattice and the prior models need to be 
defined for such a situation. Again the profile Markov random field prior is already formulated in such a 
situation. The Markov mesh formulation used here can also be extended to a three dimensional lattice, but 
the effectiveness of such a formulation need to be further studied.

 \section{Conclusion}
\label{sec:closing}
 We have compared the effect of a Markov mesh prior and a Markov random field prior, where lateral connectivity is included apriori, to predict lithology/fluid classes in a North Sea case study in a Bayesian inversion framework. The choice of prior is observed to have little influence on the posterior marginal probabilties in the current study. We have observed that the Markov mesh prior had quite a strong influence on posterior connectivity of oil sand, and was better able to capture curvature and lateral connectivity.

\section{Acknowledgments}
The authors acknowledge the Uncertainty in Reservoir Evaluation (URE) activity at the Norwegian University of Science and Technology (NTNU).

\bibliographystyle{jasa}
\bibliography{mybib}

\begin{thebibliography}{35}
\newcommand{\enquote}[1]{``#1''}
\expandafter\ifx\csname natexlab\endcsname\relax\def\natexlab#1{#1}\fi

\bibitem[\protect\citename{Abend et~al., }1965]{art133}
Abend, K., Harley, T., and Kanal, L. (1965).
\newblock \enquote{Classification of binary random patterns.}
\newblock {\em IEEE Transactions on Information Theory\/}, 11, 538--544.

\bibitem[\protect\citename{Aki and Richards, }1980]{book45}
Aki, K. and Richards, P.~G. (1980).
\newblock {\em Quantitative Seismology: Theory and Methods\/}.
\newblock New York: W. H. Freeman and Co.

\bibitem[\protect\citename{Arnesen and Tjelmeland, }2017]{art157}
Arnesen, P. and Tjelmeland, H. (2017).
\newblock \enquote{Prior specification of neighbourhood and interaction
  structure in binary {M}arkov random fields.}
\newblock {\em Statistics and Computing\/}, 27, 737--756.

\bibitem[\protect\citename{Aster et~al., }2011]{book41}
Aster, R., Borchers, B., and Thurber, C.~H. (2011).
\newblock {\em Parameter estimation and inverse problems\/}.
\newblock Amsterdam: Elsevier.

\bibitem[\protect\citename{Buland and Omre, }2003]{art162}
Buland, A. and Omre, H. (2003).
\newblock \enquote{Bayesian linearized {AVO} invserion.}
\newblock {\em Geophysics\/}, 68, 185--198.

\bibitem[\protect\citename{Connolly and Hughes, }2016]{art176}
Connolly, P. and Hughes, M. (2016).
\newblock \enquote{Stochastic inversion by matching to large numbers of
  pseudo-wells.}
\newblock {\em Geophysics\/}, 81, 2, M7--M22.

\bibitem[\protect\citename{Cressie and Davidson, }1998]{art119}
Cressie, N. and Davidson, J. (1998).
\newblock \enquote{Image analysis with partially ordered {M}arkov models.}
\newblock {\em Computational Statistics and Data Analysis\/}, 29, 1--26.

\bibitem[\protect\citename{Eidsvik et~al., }2004]{art172}
Eidsvik, J., Mukerji, T., and Switzer, P. (2004).
\newblock \enquote{Estimation of geological attributes from a well log: {A}n
  application of hidden {M}arkov chains.}
\newblock {\em Mathematical Geology\/}, 36, 379--397.

\bibitem[\protect\citename{Emery and Lantu\'{e}joul, }2014]{art165}
Emery, X. and Lantu\'{e}joul, C. (2014).
\newblock \enquote{Can a training image be a substitute for a random field
  model?}
\newblock {\em Mathematical Geosciences\/}, 46, 133--147.

\bibitem[\protect\citename{Fjeldstad and Omre, }2017]{tech34}
Fjeldstad, T. and Omre, H. (2017).
\newblock \enquote{Bayesian inversion of convolved hidden {M}arkov models with
  applications in reservoir prediction.}
\newblock Tech. rep., ArXiv e-prints 1710.06613v1, Available from
  http://arxiv.org/abs/1710.06613v1.

\bibitem[\protect\citename{Gamerman and Lopes, }2006]{book30}
Gamerman, D. and Lopes, H.~F. (2006).
\newblock {\em {M}arkov chain {M}onte {C}arlo: Stochastic simulation for
  {B}ayesian inference\/}.
\newblock 2nd ed. London: Chapman \& Hall/CRC.

\bibitem[\protect\citename{Gilks et~al., }1996]{book17}
Gilks, W.~R., Richardson, S., and Spiegelhalter, D.~J. (1996).
\newblock {\em Markov chain {M}onte {C}arlo in practice\/}.
\newblock London: Chapman \& Hall.

\bibitem[\protect\citename{Grabisch et~al., }2000]{art138}
Grabisch, M., Marichal, J.~L., and Roubens, M. (2000).
\newblock \enquote{Equivalent representations of set functions.}
\newblock {\em Mathematics of Operations Research\/}, 25, 157--178.

\bibitem[\protect\citename{Grana and {Della Rossa}, }2010]{art174}
Grana, D. and {Della Rossa}, E. (2010).
\newblock \enquote{Probabilistic petrophysical-properties estimation
  integrating statistical rock physics with seismic inversion.}
\newblock {\em Geophysics\/}, 75, 3, O21--O37.

\bibitem[\protect\citename{Grana et~al., }2017]{art164}
Grana, D., Fjelstad, T., and Omre, H. (2017).
\newblock \enquote{Bayesian {G}aussian mixture linear inversion for geophysical
  inverse problems.}
\newblock {\em Mathematical Geosciences\/}, 49, 493--515.

\bibitem[\protect\citename{Guardiano and Srivastava, }1993]{pro23}
Guardiano, F. and Srivastava, R. (1993).
\newblock \enquote{Multivariate geostatistics: beyond bivariate moments.}
\newblock In {\em Geostatistics {T}r\'{o}ia '92\/}, ed. A.~Soares,  133--144.
  Kluwer, Dordrecht.
\newblock Proceedings to the 1992 International Geostatistics Congress,
  Tr\'{o}ia, Portugal.

\bibitem[\protect\citename{Gunning and Glinsky, }2007]{art163}
Gunning, J. and Glinsky, M.~E. (2007).
\newblock \enquote{Detection of reservoir quality using Baysian seismic
  inversion.}
\newblock {\em Geophysics\/}, 72, R37--R39.

\bibitem[\protect\citename{Hammer and Holzman, }1992]{art137}
Hammer, P.~L. and Holzman, R. (1992).
\newblock \enquote{Approximations of pseudo-{B}oolean functions; applications
  to game theory.}
\newblock {\em Methods and Models of Operation Research\/}, 36, 3--21.

\bibitem[\protect\citename{Hurn et~al., }2003]{col6}
Hurn, M., Husby, O., and Rue, H. (2003).
\newblock \enquote{A tutorial on image analysis.}
\newblock In {\em Spatial statistics and computational methods\/}, ed.
  J.~M{\o}ller, vol. 173 of {\em Lecture Notes in Statistics\/},  87--139.
  Springer.

\bibitem[\protect\citename{Journel and Zhang, }2006]{art125}
Journel, J. and Zhang, T. (2006).
\newblock \enquote{The necessity of a multiple-point prior model.}
\newblock {\em Mathematical Geology\/}, 38, 591--610.

\bibitem[\protect\citename{Kindermann and Snell, }1980]{book28}
Kindermann, R. and Snell, J.~L. (1980).
\newblock {\em Markov random fields and their applications\/}.
\newblock Providence, R.I.: American Mathematical Society.

\bibitem[\protect\citename{Lang and Grana, }2017]{art173}
Lang, X. and Grana, D. (2017).
\newblock \enquote{Geostatistical inversion of prestack seismic data for the
  joint estimation of facies and impedances using stochastic sampling from
  Gaussian mixture posterior distribution.}
\newblock {\em Geophysics\/}, 82, M55--M65.

\bibitem[\protect\citename{Larsen et~al., }2006]{art169}
Larsen, A.~L., Ulvmoen, M., Omre, H., and Buland, A. (2006).
\newblock \enquote{Bayesian lithology/fluid prediction and simulation on the
  basis of a Markov-chain prior model.}
\newblock {\em Geophysics\/}, 71, R69--R78.

\bibitem[\protect\citename{Luo and Tjelmeland, }2017]{tech35}
Luo, X. and Tjelmeland, H. (2017).
\newblock \enquote{Prior specification for binary {M}arkov mesh models.}
\newblock Tech. rep., ArXiv e-prints 1707.08339v1, Available from
  http://arxiv.org/abs/1707.08339v1.

\bibitem[\protect\citename{Mariethoz and Caers, }2014]{book44}
Mariethoz, G. and Caers, J. (2014).
\newblock {\em Multiple-point geostatistics: Stochastic modeling with training
  images\/}.
\newblock John Wiley \& Sons.

\bibitem[\protect\citename{Rimstad et~al., }2012]{art146}
Rimstad, K., Avseth, P., and Omre, H. (2012).
\newblock \enquote{Hierarchical {B}ayesian lithology/fluid prediction: A
  {N}orth {S}ea case study.}
\newblock {\em Geophysics\/}, 77, B69--B856.

\bibitem[\protect\citename{Rimstad and Omre, }2010]{art171}
Rimstad, K. and Omre, H. (2010).
\newblock \enquote{Impact of rock-physics depth trends and {M}arkov random
  fields on hierarchical {B}ayesian lithology/fluid prediction.}
\newblock {\em Geophysics\/}, 75, R93--R108.

\bibitem[\protect\citename{Robert and Casella, }1999]{book24}
Robert, C.~P. and Casella, G. (1999).
\newblock {\em Monte {C}arlo statistical methods\/}.
\newblock Berlin: Springer.

\bibitem[\protect\citename{Sen and Stoffa, }2013]{book42}
Sen, M.~K. and Stoffa, P.~L. (2013).
\newblock {\em Global optimization methods in geophysical inversion\/}.
\newblock Cambridge: Cambridge University Press.

\bibitem[\protect\citename{Stien and Kolbj{\o}rnsen, }2011]{art175}
Stien, M. and Kolbj{\o}rnsen, O. (2011).
\newblock \enquote{Facies Modeling Using a Markov Mesh Model Specification.}
\newblock {\em Mathematical Geosciences\/}, 43, 6, 611.

\bibitem[\protect\citename{Strebelle, }2002]{art121}
Strebelle, S. (2002).
\newblock \enquote{Conditional simulation of complex geolgical structures using
  multiple-point statistics.}
\newblock {\em Mathematical Geology\/}, 34, 1--21.

\bibitem[\protect\citename{Tarantola, }2005]{book43}
Tarantola, A. (2005).
\newblock {\em Inverse problem theory\/}.
\newblock Philidelphia: SIAM.

\bibitem[\protect\citename{Toftaker and Tjelmeland, }2013]{art168}
Toftaker, H. and Tjelmeland, H. (2013).
\newblock \enquote{Construction of binary multi-grid {M}arkov random field
  prior models from training images.}
\newblock {\em Mathematical Geosciences\/}, 45, 383--409.

\bibitem[\protect\citename{Ulvmoen and Omre, }2010]{art123}
Ulvmoen, M. and Omre, H. (2010).
\newblock \enquote{Improved resolution in {B}ayesian lithology/fluid inversion
  from prestack seismic data and well observations: Part 1-Methodology.}
\newblock {\em Geophysics\/}, 75, R21--R35.

\bibitem[\protect\citename{Zhang et~al., }2012]{art167}
Zhang, T., Pedersen, S.~I., Knudby, C., and McCormick, D. (2012).
\newblock \enquote{Memory-efficient categorical multi-point statistics
  algorithms based on compact search trees.}
\newblock {\em Mathematical Geosciences\/}, 44, 863--879.

\end{thebibliography}

\appendix

\section{Fitted Markov mesh prior, $p(\kappa)$}
\label{app:prior}
The Markov mesh model fitted to the training image in Figure \ref{fig:training} has 
template sequential neighborhood $\tau = \{(-1,0), (0,-1), (-1,2), (0,-2), (-3,-1), (0,-3)$, 
$(-1,4), (0,-4), (-2,-4)\}$ 
and $\Delta$ and $\{\beta(\lambda):\lambda\in\Lambda\}$ are as specified in Table \ref{tab:prior}.
\begin{table}
\caption{\label{tab:prior}The elements $\lambda$ in the set $\Lambda$ and the associated interaction parameters $\beta(\lambda)$.}
\begin{center}
\begin{tabular}{c|c}            %
$\lambda\in\Lambda$ & $\beta(\lambda)$ \\ \hline
$\emptyset$ & $-4.33884$ \\
$\{(-1,0)\}$ & $3.27479$\\
$\{(0,-1)\}$ & $2.96595$ \\
$\{ (-1,0),(0,-1)\}$ & $-0.460735$\\
$\{ (-1,2)\}$ & $1.49237$ \\
$\{ (-1,2),(0,-1)\}$ & $-1.10759$\\
$\{ (0,-2)\}$ & $1.99035$\\
$\{ (-3,-1)\}$ & $-1.43573$\\
$\{ (0,-3)\}$ & $3.06786$ \\
$\{ (-1,0),(0,-3)\}$ & $-3.44258$\\
$\{ (0,-3),(0,-1)\}$ & $-2.03335$\\
$\{ (-1,0),(0,-3),(0,-1)\}$ & $1.95605$\\
$\{ (0,-3),(0,-2)\}$ & $-1.02729$\\
$\{ (-1,4)\}$ & $2.90431$\\
$\{ (-1,0),(-1,4)\}$ & $-3.42674$\\
$\{ (-1,4),(0,-1)\}$ & $-0.404195$\\
$\{ (-1,2),(-1,4)\}$ & $0.268767$\\ %
$\{ (-1,4),(0,-3)\}$ & $-2.73426$\\
$\{ (-1,0),(-1,4),(0,-3)\}$ & $2.96929$\\
$\{ (-1,4),(0,-3),(0,-1)\}$ & $1.95346$\\
$\{ (0,-4)\}$ & $2.1858$\\
$\{ (-1,0),(0,-4)\}$ & $-0.355664$\\
$\{ (0,-4),(0,-2)\}$ & $-1.61185$\\
$\{ (0,-4),(0,-3)\}$ & $-1.23267$\\
$\{ (-1,0),(0,-4),(0,-3)\}$ & $0.606075$\\
$\{ (0,-4),(0,-3),(0,-2)\}$ & $2.03717$\\
$\{ (-1,4),(0,-4)\}$ & $-4.01512$\\
$\{ (-1,0),(-1,4),(0,-4)\}$ & $3.80173$\\
$\{ (-1,4),(0,-4),(0,-3)\}$ & $2.6053$\\
$\{ (-1,0),(-1,4),(0,-4),(0,-3)\}$ & $-1.64379$\\
$\{ (-2,-4)\}$ & $-0.717159$
\end{tabular}
\end{center}
\end{table}
\end{document}